\documentclass[traditabstract]{aa}        
\usepackage{graphicx}
\usepackage{times,epsfig,amssymb}
\usepackage{txfonts}
\usepackage{natbib}
\def\ltsima{$\; \buildrel < \over \sim \;$}
\def\simlt{\lower.5ex\hbox{\ltsima}}
\def\gtsima{$\; \buildrel > \over \sim \;$}
\def\simgt{\lower.5ex\hbox{\gtsima}}
\begin{document}
   \title{Metal Abundances in the Cool-Cores of Galaxy Clusters}


   \author{Sabrina De Grandi
           \inst{1}
           \and
           Silvano Molendi
           \inst{2}
           }

   \institute{INAF - Osservatorio Astronomico di Brera,
              via E. Bianchi 46, I-23807 Merate (LC)\\
              \email{sabrina.degrandi@brera.inaf.it}
         \and
              INAF - IASF Milano
               via Bassini 15, I-20133 Milano \\
             \email{silvano@iasf-milano.inaf.it}
             }

   \date{Received ; accepted }

   \abstract {We use {\it XMM-Newton} data to carry out a detailed
   study of the Si, Fe and Ni abundances in the cool cores of a
   representative sample of 26 local clusters.  We have performed a
   careful evaluation of the systematic uncertainties related to the
   instruments, the plasma codes and the spectral modeling finding
   that the major source of uncertainty is in the plasma codes.  Our
   Si, Fe, Ni, Si/Fe and Ni/Fe distributions feature only moderate
   spreads (from 20\% to 30\%) around their mean values strongly
   suggesting similar enrichment processes at work in all our cluster
   cores.  Our sample averaged Si/Fe ratio is comparable to those
   measured in samples of groups and high luminosity ellipticals
   implying that the enrichment process in ellipticals, dominant
   galaxies in groups and BCGs in clusters is quite similar.  Although
   our Si/Fe and Ni/Fe abundance ratios are fairly well constrained,
   the large uncertainties in the supernovae yields prevent us from
   making a firm assessment of the relative contribution of type Ia
   and core-collapsed supernovae to the enrichment process.  All that
   can really be said with some certainty is that both contribute to
   the enrichment of cluster cores.

}

   \titlerunning{Metal Abundances in the Cool-Cores of Galaxy Clusters}
   \authorrunning{S. De Grandi \& S. Molendi}
   \maketitle

\section{Introduction}

The intra-cluster medium (ICM), that fills the deep potential well of
galaxy clusters, is rich in metals. In the high mass regime of rich
clusters ($M_{vir}\gtrsim 10^{14} M_\odot$), where temperatures and
densities of the gas are of the order of $3-10$ keV and $10^{-3}$
cm$^{-3}$, respectively, the ICM is heavily ionized and the main
processes producing emission lines are collisions of ions with
electrons in a plasma in collisional ionization equilibrium.  Since
the elements are well confined within the cluster potential well they
accumulate over the whole cluster history and retain important
information on cluster formation and evolution.  For gas temperatures
$\gtrsim 3$ keV the most prominent emission lines are from He-like and
H-like K-shell transitions of Fe (at $\sim 7$ keV), along with less
pronounced K-shell lines of S, Si (at $\sim 2$ keV), Ar, Ca and Ni (at
$\sim 8$ keV).
Temperature and abundances of the ICM are measured at the same time
from the X-ray spectrum: the temperature is measured from the
continuum emission that is almost entirely given by thermal
bremsstrahlung, whereas the abundance of an element is derived by
measuring the equivalent width of the line (once the continuum is
known), that is directly proportional to the ion-to-Hydrogen
concentrations ratio \citep[see][for a review of the thermal radiation
processes in clusters]{kaastra08_rev}.
While in principle the abundance measurement is straightforward, in
practice various sources of uncertainties are present, amongst which
(1) the accuracy of the atomic physics, (2) the moderate spectral
resolution of the current imaging instruments which often results in
line blending, this is particularly severe for L-shell blends where
the transitions are not entirely understood, (3) the presence of
temperature gradients in the ICM, especially in the cluster cores,
that need specific spectral modeling.
With the advent of {\it Chandra} and {\it XMM-Newton} satellites,
carrying detectors with both high spatial and spectral capabilities,
the statistical quality of cluster spectra (in particular for
cool-core regions) improved dramatically.  While on the one hand the
statistical errors associated with the abundance measurements have
greatly decreased, on the other surprisingly little attention has been
devoted to systematic sources of uncertainties which, under these
circumstances, are likely to play an important role.

In this paper our first goal is to provide a robust measurement of the
distribution of abundances and abundance ratios of the chemical
elements in the cores of a representative sample of nearby and bright
cool-core clusters.  In this context ``robust estimate'' essentially
means that we will include in the error budget a careful evaluation of
the systematic uncertainties potentially affecting our data.  We will
consider both instrument related systematics and plasma code
systematics. We stress that in this paper we are interested in the
measure of the global abundances in the central region of each
cluster, not in the analysis of radial profiles. Moreover these
central regions will provide us with the maximum photon statistics
because of their very intense central surface brightness peaks,
thereby allowing us to explore the most subtle sources of systematic
errors and reliably measure the most abundant elements observable in
the ICM (e.g. Si, Fe and Ni).

The elements observed in the ICM are produced through thermo-nuclear
nucleo-synthesis in supernovae (SN) explosions occurring in the member
early-type galaxies \citep{arnaud92,renzini93}, and they are
eventually ejected/diffused into the ICM through galactic winds
\citep[for a review see][]{bland07} and ram pressure
stripping \citep[e.g.][and references therein]{schindler08_rev}.
Supernovae are classified on the basis of their progenitor
models. Nowaday two main groups of SNe are known: type Ia supernovae
(SNIa) that derive from an accreting white dwarf in a binary system,
and, core-collapsed supernovae (SNcc) whose progenitors are single
massive stars ($\gtrsim 8 M_\odot$).  One of the most important
quantity that theoretical models of SNe provide are the yields of the
various chemical elements, namely the mass per element and per SN
event. While, SNIa produce mostly Fe, Ni and Si-group metals (i.e.
Si, S, Ar, Ca), the SNcc ejecta are rich in $\alpha$-elements (i.e. O,
Ne, Mg, as well as few Si-group elements).
In the 90's {\it BeppoSAX} and {\it ASCA} observations revealed that
the centers of cool-core clusters always display an excess (with
respect to the external cluster regions) of Fe around the central
cluster galaxy
\citep[e.g.][]{fukazawa00,degrandi01_ab,leccardi08_ab}. The iron mass
inferred from this excess is consistent with being entirely produced
by the giant galaxy itself \citep[e.g.][]{degrandi04, boehringer04b},
which is always found in these systems. Other elements such as Mg, Si
and S show abundance peaks in cluster cores
\citep[e.g.][]{fukazawa98,finoguenov00,finoguenov01,tamura04,sato09_a262}.
Under the assumption that the sole source of metals are the two types
of SNe, observations of the $\alpha$-elements/Fe abundance ratios
coupled with the knowledge of the SNe chemical yields allow the
determination of the SNIa and SNcc proportion within each cluster.
\citet{finoguenov00} found an increasing Si/Fe ratio with radius
in clusters indicating a greater predominance of SNcc enrichment at
large radii, while the innermost parts appeared dominated by SNIa
products. However works based on data taken with the more recent {\it
XMM-Newton} and {\it SUZAKU} observatories have not fully confirmed
this overall picture for rich clusters.  \citet{tamura04} studying the
spatial distributions of metals in a sample of cool-core clusters
observed with {\it XMM-Newton} found uniform Si/Fe and S/Fe ratios
within the ICM, but increasing O/Fe (although this ratio was prone to
large uncertainties). {\it SUZAKU} observations confirm these findings
\citep[e.g.][]{sato07_sn,matsushita08}.
The relative proportion of SNIa and SNcc found by {\it XMM-Newton} and
{\it SUZAKU} observations are in raw agreement \citep{sato07_sn},
although these results were achieved under the choice of specific
compilations of SNe yields \citep[e.g.][]{iwamoto99,nomoto06}.  It is
worth noting that none of the combinations of the theoretical SNIa and
SNcc yields available in the literature have reproduced the overall
elemental pattern in cluster cores \citep[fits were not formally
acceptable based on the $\chi^2$ values; e.g.][]
{baumgartner05,deplaa07,sato07_sn}.  There are at least two possible
solutions to this problem: the existence of an additional source of
elements other than SNIa and SNcc \citep[see][]{finoguenov02,
baumgartner05,mannucci06}, or, the need of a revision of the
theoretical SNe models \citep[e.g.][]{young07}.
Interestingly a systematic exploration of the full range of the
uncertainties of theoretical yields was last provided by
\citet{gibson97}, although since then, several new compilations of
SNcc yields have become available \citep[e.g.][]{chieffi04,nomoto06}
along with new yields for SNIa \citep{iwamoto99}.  As pointed out by
\cite{gibson97} the relative contribution of SNIa and SNcc to the Fe
abundance in the ICM depends crucially upon the adopted theoretical
SNe models.

The second main goal of this paper is to provide a critical assessment
of the relative contributions of the SNIa versus SNcc to the
enrichment process.  Contrary to other similar works present in the
recent literature \citep[e.g.][]{deplaa07} we do not neglect the
uncertainties associated to the current theoretical yields and explore
how they affect the derived SNIa fraction.

This paper is organized as follows. In Sect. 2 we present the sample,
in Sect. 3 we describe the cleaning process of the raw {\it
XMM-Newton} archival data, and, in Sect. 4, we concentrate on the
spectral analysis and on the choice of the best fitting model. In
Sect. 4 we also discuss the systematic uncertainties related to
imperfections in the cross-calibration between the three EPIC
instruments. In Sect. 5 we present results on the derived
distributions of metal abundances and abundances ratios. In this same
section we compare results from two different choices for the spectral
codes. In Sect. 6 we discuss the relative contribution to the overall
enrichment process of different SNe types. In Sect. 7 we discuss our
findings and in Sect. 8 we summarize our main results.

We assume $H_0 = 70$ km s$^{-1}$ Mpc$^{-1}$ and
$\Omega_{\Lambda}=0.7$, the source redshifts are all extracted from
the NASA/IPAC Extragalactic Database (NED).  All metallicity
measurements we show in this paper are relative to the photospheric
solar abundances of \citet{anders89}.  In this set of abundances, Fe
has a number density of $4.68\times 10^{-5}$, Si $3.55\times 10^{-5}$
and Ni $1.78\times 10^{-6}$, all number densities are relative to
H. All uncertainties shown are $1\sigma$ confidence level (i.e
$68\%$).

\section{The sample}

Starting from the flux-limited sample of 55 galaxy clusters listed by
\citet{edge92} and the ROSAT PSPC analysis of this sample
performed by \citet{peres98}, we have selected all the
objects with a mass deposition rate different from zero
\citep[see Table 5 in][]{peres98}.
Among these clusters we have chosen all those that were observed with
{\it XMM-Newton} EPIC within April 2008, with the exclusion of
clusters whose observations are strongly affected by soft protons
(e.g. Abell 644 and Abell 2142).  The 32 clusters that meet these
requirements are listed in Table ~\ref{tab_sample}.

 \begin{table*}
 \begin{center}
 \caption{Starting sample of 32 galaxy clusters extracted from the X-ray
 flux limited B55 sample \citep{edge92}. All objects are cool-core
 clusters with a mass deposition rate different from zero \citep{peres98}.}
 \label{tab_sample}
 \begin{tabular}{l@{\hspace{.8em}} c@{\hspace{.6em}} 
c@{\hspace{.8em}} c@{\hspace{.8em}} c@{\hspace{.8em}} 
c@{\hspace{.8em}} r@{\hspace{.8em}}  c@{\hspace{.8em}}} \hline \hline
Cluster & Obs ID & $z^{\mathrm a}$ & $N_H^{\mathrm b}$          &      & Exp.time$^{\mathrm c}$ &        & $r_{cool}/2$$^{\mathrm d}$ \\
        &        &                 & ${\mathrm 10^{22}cm^{-2}}$ & MOS1 & MOS2                   & {\it pn}& arcmin (kpc) \\
\hline
Abell 85         & 0065140101 & 0.0551 & 0.0342 & 12.5 & 12.5 &  9.7 & 1.137 (73)  \\
Abell 262        & 0109980101 & 0.0163 & 0.0538 & 23.8 & 23.9 & 19.2 & 2.613 (52)  \\
AWM7             & 0135950301 & 0.0172 & 0.0983 & 30.7 & 30.7 & 28.7 & 2.473 (52)  \\
Abell 3112       & 0105660101 & 0.0750 & 0.0261 & 23.3 & 23.3 & 18.6 & 1.124 (96)  \\
Perseus          & 0085110101 & 0.0179 & 0.1410 & 53.7 & 53.7 & 51.3 & 4.263 (93)  \\ 
2A $0335+096$    & 0147800201 & 0.0349 & 0.1780 & 80.2 & 80.9 & 73.9 & ~~2.596 (108) \\
Abell 478        & 0109880101 & 0.0881 & 0.1510 & 28.7 & 79.0 & 42.6 & ~~1.032 (102) \\
Abell 496        & 0135120201 & 0.0329 & 0.0458 & 17.8 & 17.6 & 14.1 & 1.397 (55)  \\
PKS $0745-191$   & 0105870101 & 0.1028 & 0.4240 & 18.9 & 18.9 &  9.8 & ~~0.944 (107) \\
Hydra A          & 0109980301 & 0.0549 & 0.0494 & 18.6 & 19.8 & 15.8 & 1.266 (81)  \\
Abell 1060       & 0206230101 & 0.0126 & 0.0490 & 40.0 & 41.5 & 33.2 & 2.588 (40)  \\
Virgo (M87)      & 0200920101 & 0.0044 & 0.0254 & 77.8 & 79.4 & 70.0 & 9.352 (51)  \\
Centaurus        & 0046340101 & 0.0114 & 0.0806 & 40.7 & 41.3 & 40.3 & 2.928 (41)  \\
Abell 1644       & 0010420201 & 0.0473 & 0.0499 & 15.1 & 15.2 & 12.8 & 0.521 (29)  \\
Abell 1650       & 0093200101 & 0.0838 & 0.0156 & 38.1 & 36.6 & 34.4 & 0.878 (83)  \\
Abell 1651       & 0203020101 & 0.0849 & 0.0181 & 10.2 & 10.8 &  6.9 & 0.669 (64)  \\
Abell 1689       & 0093030101 & 0.1830 & 0.0182 & 37.5 & 37.6 & 32.6 & 0.520 (96)  \\
Abell 3558       & 0107260101 & 0.0480 & 0.0389 & 44.1 & 44.0 & 38.9 & 0.602 (34)  \\
Abell 3562       & 0105261301 & 0.0490 & 0.0385 & 40.3 & 40.2 & 38.0 & 0.834 (48)  \\
Abell 3571       & 0086950201 & 0.0391 & 0.0371 & 24.3 & 24.5 & 16.5 & 1.119 (52)  \\
Abell 1795       & 0097820101 & 0.0625 & 0.0119 & 39.5 & 39.6 & 22.7 & 1.232 (89)  \\
Abell 2029       & 0111270201 & 0.0773 & 0.0304 & 10.7 & 11.2 & 10.3 & 1.059 (93)  \\
Abell 2052       & 0109920101 & 0.0355 & 0.0272 & 30.3 & 30.3 & 26.5 & 1.747 (74)  \\
MKW3s            & 0109930101 & 0.0450 & 0.0303 & 38.7 & 38.3 & 34.1 & 1.620 (86)  \\
Abell 2065       & 0202080201 & 0.0726 & 0.0295 & 20.7 & 20.7 & 16.6 & 0.338 (28)  \\
Abell 2063       & 0200120401 & 0.0349 & 0.0298 &  8.1 &  8.9 &  5.4 & 1.152 (48)  \\
Abell 2199       & 0008030201 & 0.0302 & 0.0086 & 14.6 & 14.6 & 10.0 & 1.985 (72)  \\
Abell 2204       & 0112230301 & 0.1523 & 0.0567 & 19.9 & 20.2 & 13.7 & ~~0.629 (100) \\
Cygnus A         & 0302800101 & 0.0561 & 0.3490 & 22.2 & 22.2 & 20.2 & 1.041 (68)  \\
Abell 2597       & 0147330101 & 0.0852 & 0.0249 & 52.2 & 53.3 & 48.5 & 0.792 (76)  \\
Abell 4038       & 0204460101 & 0.0300 & 0.0155 & 29.4 & 29.2 & 27.6 & 1.859 (67)  \\
Abell 4059$^{e}$ & 0109950101 & 0.0475 & 0.0110 & 12.7 & 13.7 &  7.0 & 1.378 (77)  \\
Abell 4059       & 0109950201 & 0.0475 & 0.0110 & 23.3 & 23.2 & 19.7 & 1.378 (77)  \\
\hline
\multicolumn{8}{l}{{\bf Notes}: $^{\mathrm a}$ redshifts taken from the NASA Extragalactic 
Database; $^{\mathrm b}$ Galactic photoelectric }\\
\multicolumn{8}{l}{absorption extracted from the HEASARCH Database (LAB survey 
\citealt{kaberla05});}\\
\multicolumn{8}{l}{$^{\mathrm c}$ net exposure times for the three EPIC instruments 
after data cleaning  in ks; }\\
\multicolumn{8}{l}{$^{\mathrm d}$ extraction radius is half the cooling radius from 
\cite{peres98}; $^{\mathrm e}$ the two } \\
\multicolumn{8}{l}{observations of Abell 4059 were merged together before the spectral 
analysis.}\\
\end{tabular}


 \end{center}
 \end{table*}

\section{Data preparation}

We reprocessed the Observation Data Files (ODF) using the Science
Analysis System (SAS) version 7.0.0.  After the production of the
calibrated event lists for the EPIC MOS1, MOS2 and {\it pn}
observations with {\it emchain} and {\it epchain} tasks, we have
performed a soft proton cleaning using a double filtering process.

We first removed soft protons spikes by screening the light curves
produced in 100 bins in the 10-12 keV band and applying an opportune
threshold for each instrument, generally a threshold of 0.20 cts
s$^{-1}$ for MOS1 and MOS2, and of 0.60 cts s$^{-1}$ for {\it
pn}. Then to eliminate possible residual flares contributing below 10
keV, we extracted a light curve in the 2-5 keV band and fitted the
histogram produced from this curve with a Gaussian distribution. To
generate the final filtered event file we have rejected all events
registered at times when the count rate was more than $3\sigma$ from
the mean of this distribution.  Finally, we have filtered events files
according to FLAG (FLAG==0) and PATTERN (PATTERN$\leq 12$ for MOS and
PATTERS==0 for {\it pn}) criteria.  The resulting effective exposure
times of the observations are reported in Table ~\ref{tab_sample}.

Using the cleaned events files we have extracted spectra from a
circular region centered on the clusters emission peaks. The physical
radius of this region is 0.5~r$_{\rm cool}$, where r$_{\rm cool}$ is
the cooling radius computed by \citet{peres98}.  With this choice of
the extraction radius we sample a significative portion of the cool
core and we assure that the extracted region is contained within the
EPIC field of view for all our clusters.  Prominent point-like sources
have been removed from the extraction region.  Other authors
\citep[e.g.][]{deplaa07,rasmussen07} select there extraction regions
as fixed fractions of the cluster scaling radius,
e.g. \cite{deplaa07} used 0.2 of r$_{500}$, where r$_{500}$ is the radius
encompassing a spherical density contrast of 500 with respect to the
critical density. By converting our extraction radii in r$_{500}$
units we find a rather peaked distribution centered around 0.08
r$_{500}$.

   \begin{figure*} \centering
   \includegraphics[width=4.5cm,angle=-90]{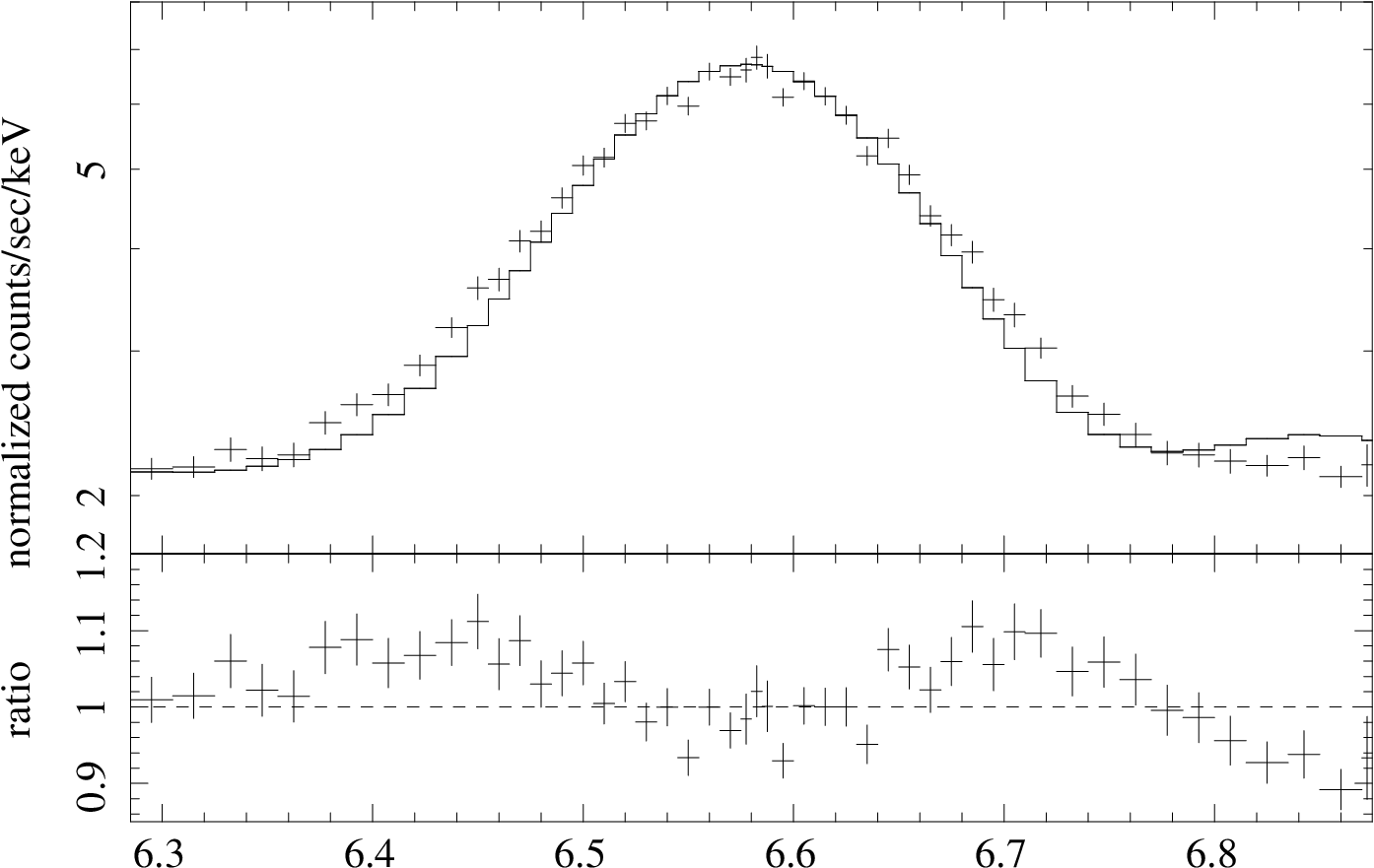}
   \includegraphics[width=4.5cm,angle=-90]{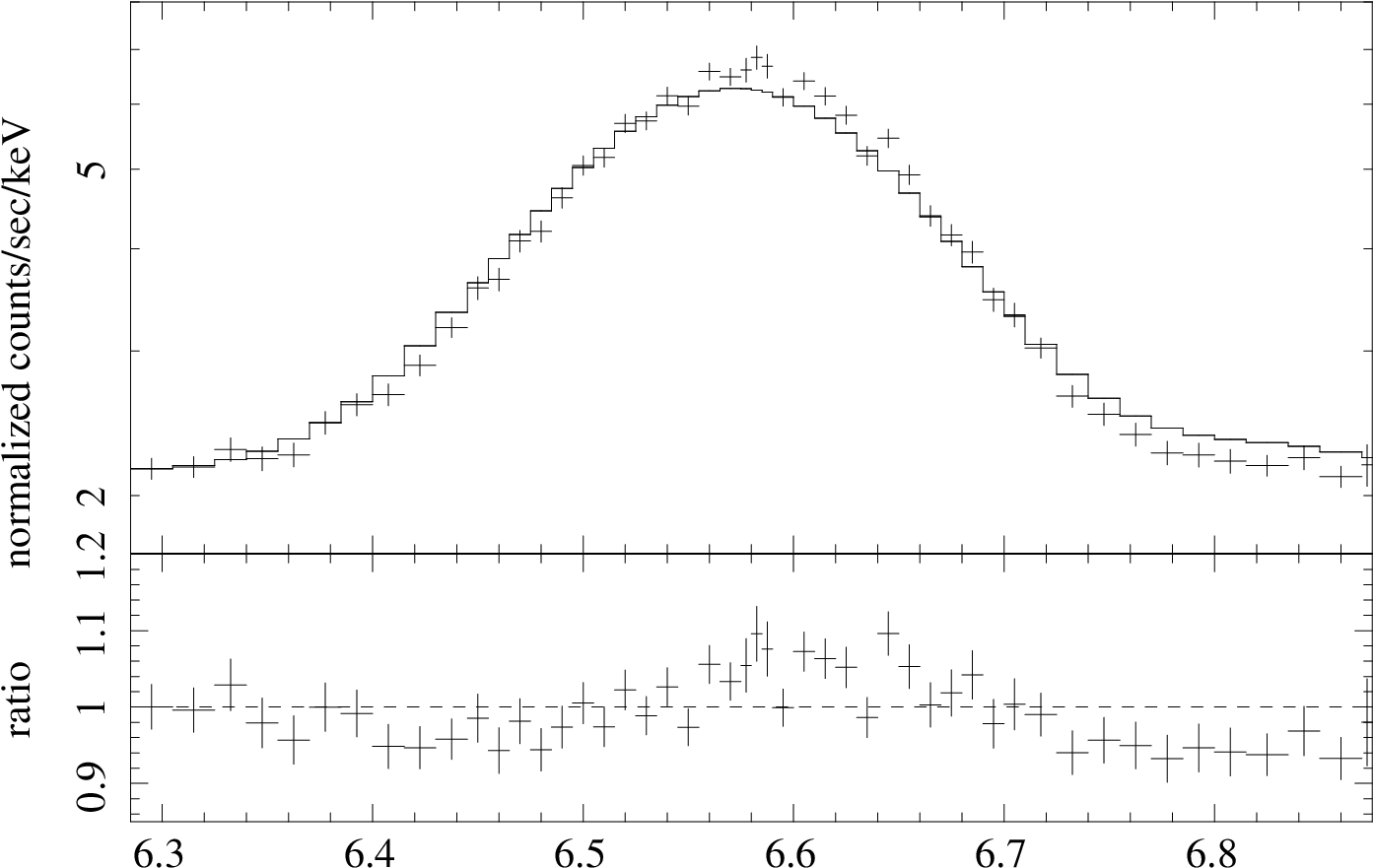}
   \includegraphics[width=4.5cm,angle=-90]{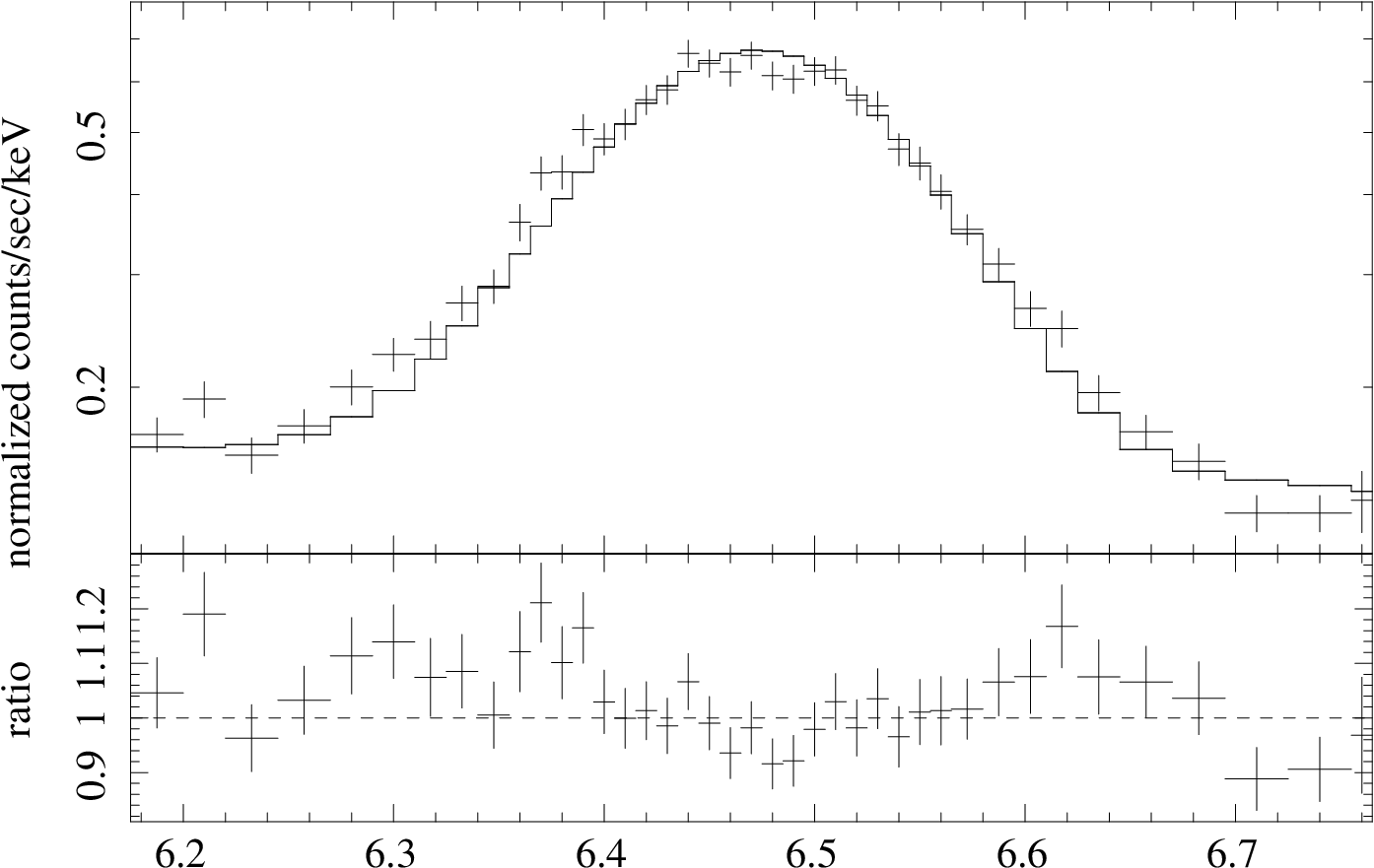}
   \includegraphics[width=4.5cm,angle=-90]{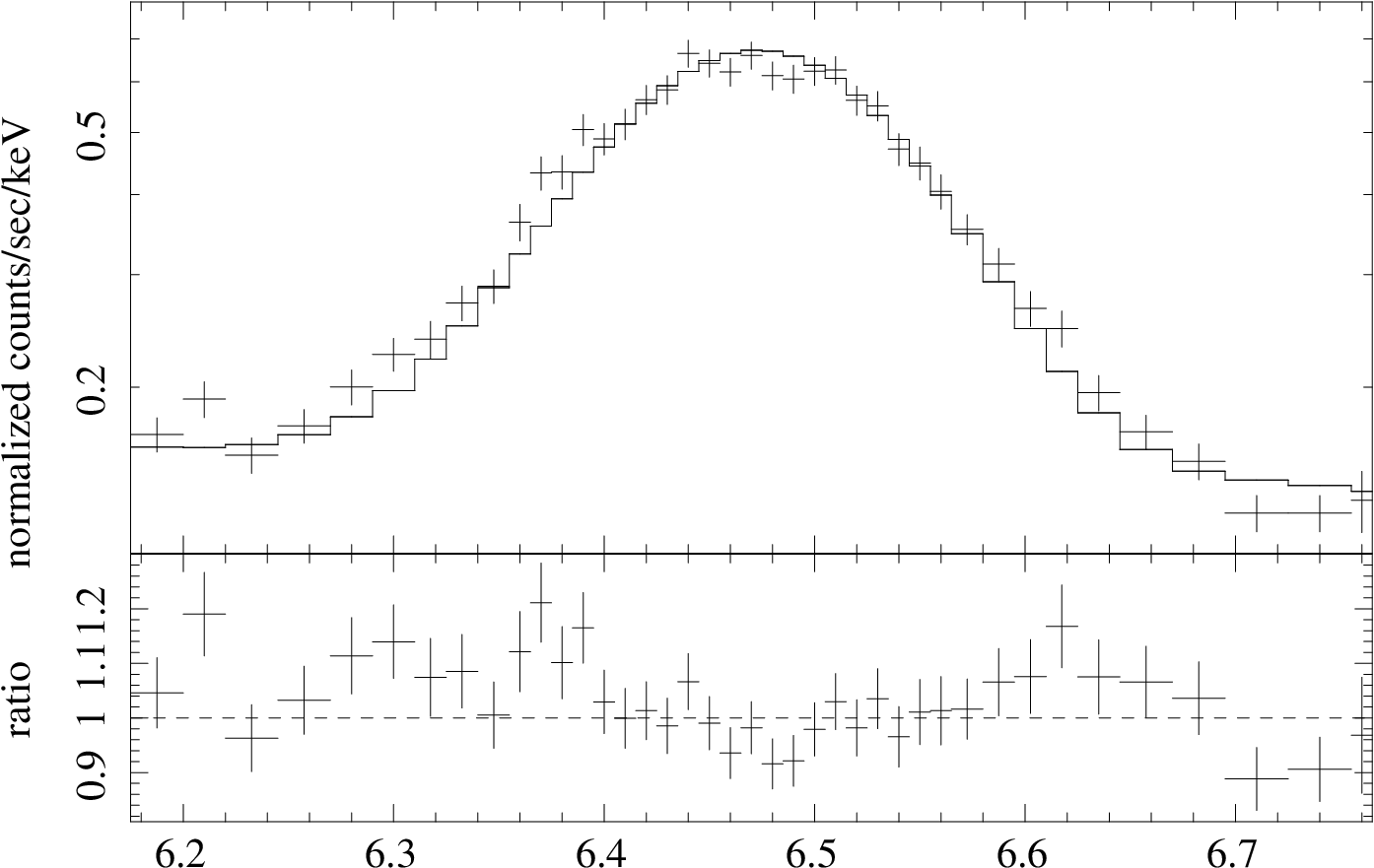}
   \includegraphics[width=4.8cm,angle=-90]{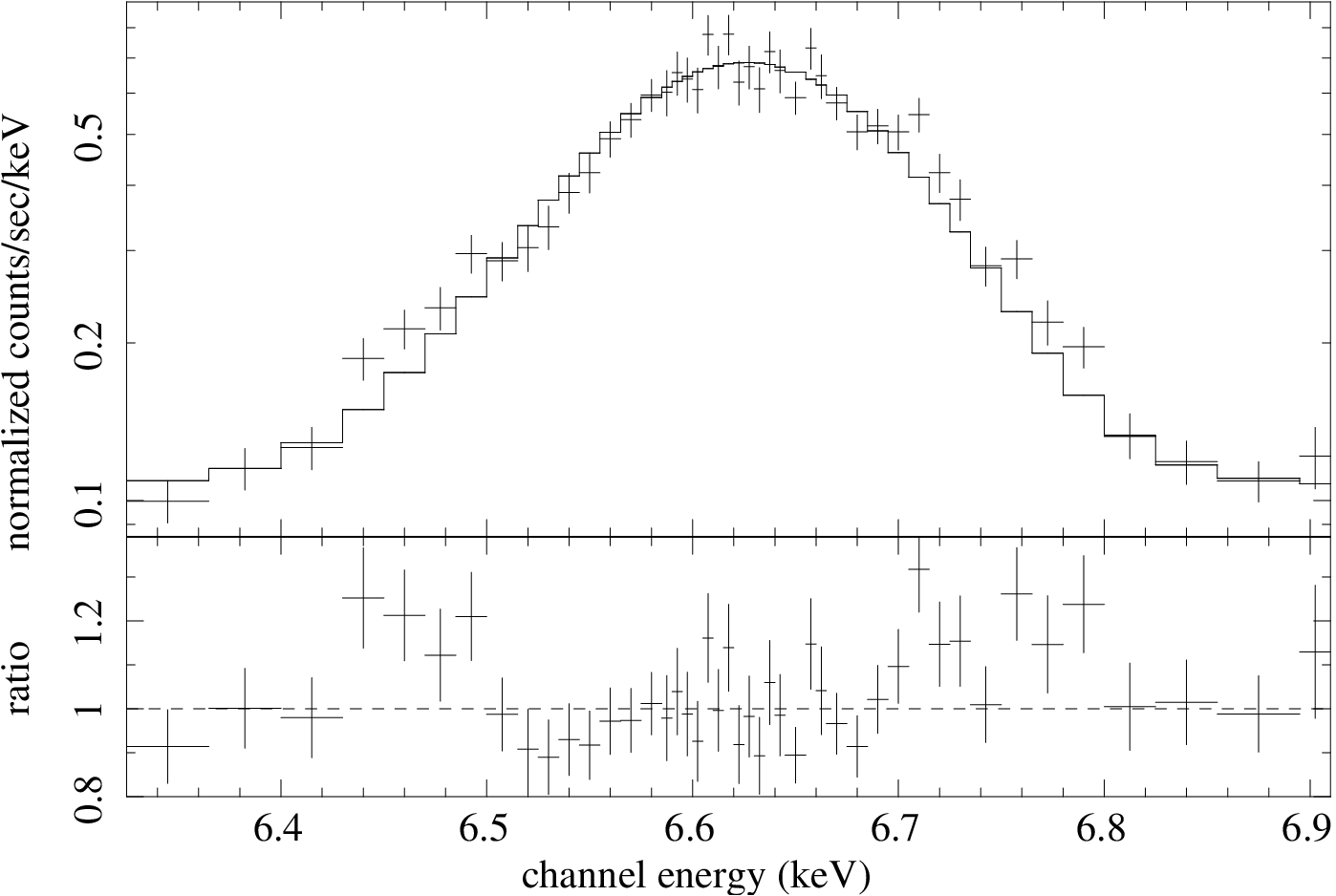}
   \includegraphics[width=4.8cm,angle=-90]{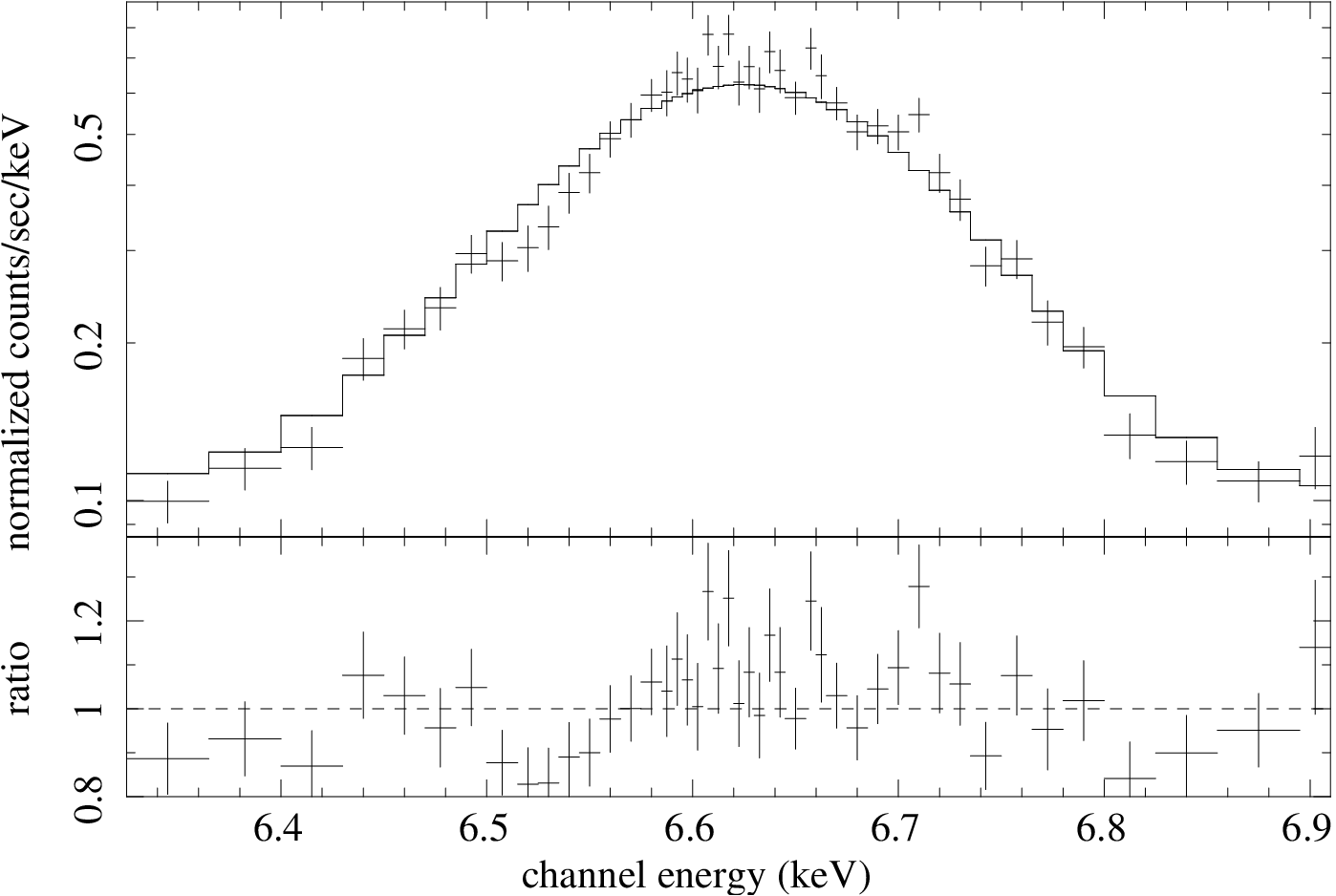}
   \caption{EPIC-{\it pn} spectral data (crosses) and best-fit models
   (continuum lines) for 3 high counting statistics clusters: Perseus
   (on the top), 2A $0335+096$ (in the middle) and Centaurus (at the
   bottom). Panels on the left show results from a 4T model fit
   without Gaussian smoothing whereas plots on the right are
   smoothed. In all cases we also plot the data divided by the
   folded model. The spectra are rebinned to allow graphical
   clarity. The energy range of each plot is 600 eV centered on the
   6.7 keV line (redshifted at the cluster distance). In the panels on
   the left are clearly visible the ``wings'' around the Fe line in
   the residuals that are an indication of the resolution
   miss-calibration of the redistribution matrix of the {\it pn} (see
   discussion in Sect. 4.1).}
   \label{fig_smo}
   \end{figure*}

 \begin{table*}
 \begin{center}
 \caption{Relative differences between metal measurements made with
 the three EPIC instruments MOS1, MOS2 and {\it pn} computed over the
 whole sample. For each element in the first line only statistical
 errors are accounted for, in the second line a $3\%$ systematic error
 is summed up in quadrature to each abundance measurement (see
 discussion in Sect. 4.3).}
 \label{tab_cross}
 \begin{tabular}{l c c c}            
& & & \\
\hline\hline                        
{\it mekal} model & & & \\
\hline
Metal $X$ & $(X_{MOS1}-X_{MOS2})/X_{MOS2}$ & $(X_{MOS1}-X_{pn})/X_{pn}$ &
$(X_{MOS2}-X_{pn})/X_{pn}$ \\       
\hline
    Fe                & $~~0.02\pm 0.01$ & $0.07\pm 0.01$ & $0.02\pm 0.02$ \\
$+3\%$ syst.err.      & $~~0.02\pm 0.02$ & $0.04\pm 0.02$ & $0.01\pm 0.02$ \\
\hline
    Si                & $~~0.01\pm 0.02$ & $0.19\pm 0.03$ & $0.10\pm 0.03$ \\
$+3\%$ syst.err.      & $ -0.03\pm 0.04$ & $0.11\pm 0.04$ & $0.08\pm 0.04$ \\
\hline
    Ni                & $ -0.12\pm 0.08$ & $0.14\pm 0.11$ & $0.01\pm 0.09$ \\
$+3\%$ syst.err.      & $ -0.13\pm 0.08$ & $0.12\pm 0.11$ & $0.00\pm 0.10$ \\
\hline
& & & \\
& & & \\
\hline\hline                        
{\it apec} model & & & \\
\hline
Metal $X$ & $(X_{MOS1}-X_{MOS2})/X_{MOS2}$ & $(X_{MOS1}-X_{pn})/X_{pn}$ &
$(X_{MOS2}-X_{pn})/X_{pn}$ \\       
\hline
    Fe                & $~~0.03\pm 0.01$ & $0.07\pm 0.01$ & $0.02\pm 0.02$ \\
$+3\%$ syst.err.      & $~~0.02\pm 0.02$ & $0.04\pm 0.02$ & $0.01\pm 0.02$ \\
\hline
    Si                & $~~0.01\pm 0.03$ & $0.22\pm 0.03$ & $0.13\pm 0.04$ \\
$+3\%$ syst.err.      & $ -0.03\pm 0.05$ & $0.13\pm 0.04$ & $0.11\pm 0.04$ \\
\hline
    Ni                & $ -0.16\pm 0.09$ & $0.13\pm 0.14$ & $0.01\pm 0.11$ \\
$+3\%$ syst.err.      & $ -0.17\pm 0.10$ & $0.12\pm 0.14$ & $0.01\pm 0.11$ \\
\hline
\end{tabular}

 \end{center}
 \end{table*}

\section{Spectral analysis}

In this paper we wish to measure abundances for Si, Fe and Ni.  To
achieve robust estimates we choose to make our measures from the
K-shell lines only, avoiding L-shell blends where uncertainties
associated to the atomic physics are larger.  We achieve this by
setting a lower energy threshold of 1.8 keV which includes Si K-shell
emission (2 keV) and excludes the Fe and Ni L-shell blends ($\sim$
1.-1.5 keV).

The background subtraction for the high surface brightness core
regions of cool-core clusters is less critical than in the case of
more external low surface brightness regions.  We therefore have
decided to subtract the background using blank-sky fields instead of
proceeding with a more detailed modeling of the different background
components.  The blank-sky fields for EPIC MOS and {\it pn} were
produced by \citet{leccardi08_ab} (see Appendix B ``The analysis
of blank field observations'' in their paper) by analysing a large
number of observations for a total exposure time of $\sim 700$ ks for
MOS and $\sim 500$ ks for {\it pn}.

We have refined our background analysis performing also a background
rescaling for each observation separately to account for temporal
variations of the background. We have estimated the background
intensity from a spectra extracted from an external ring between
$10^\prime$ and $12^\prime$ centered on the emission peak, taking into
account only the $10-12$ keV band (to avoid possible extended cluster
emission residuals in this region). We have than rescaled the
blank-sky fields background to the local value.  This procedure is
important especially when deriving the nickel abundance from its
emission lines at $\sim 8$ keV. Indeed, in this hard spectral region
both the EPIC effective area and the surface brightness of the
relatively low temperature cluster cores decrease rapidly, and, the
background becomes progressively more important.

All the spectral fits were performed with the XSPEC package (version
11.3.2, \citealt{arnaud96_xspec}).

We have analyzed each cluster spectra with three different models: (1)
a one temperature thermal model with the plasma in collisional
ionization equilibrium ({\it vmekal} model in the XSPEC nomenclature),
referred as 1T model thereafter, (2) a two temperature thermal model
({\it vmekal+vmekal}), 2T model thereafter, and, (3) a
multi-temperature model ({\it vmekal+vmekal+vmekal+vmekal}), 4T model
thereafter.

All models have been multiplied by the Galactic hydrogen column
density, $N_H$, determined by HI surveys \citep{kaberla05} through the
{\it wabs} absorption model in XSPEC. In the following sections we
report results from spectral fits with $N_H$ fixed at the weighted
Galactic value only. We have also allowed $N_H$ to vary in all three
models finding no significant differences in the derived silicon, iron
and nickel abundance values.

The redshifts have been taken from the NASA/IPAC Extragalactic
Database and have been always left as free parameters to account for
small calibration differences between the two MOS and the {\it pn}.
We have allowed the temperatures of 1T and 2T models to vary freely,
whereas in the 4T model temperatures of the four components were fixed
at 1, 2, 4 and 8 keV, respectively.  The normalizations of the models
and the elements Si, S, Fe and Ni were left free to vary.  Ar and C
were jointed to S as these elements are less abundant and we do not
plan to study them in the course of this work.

As pointed out by \citet{leccardi08_ab}, when fitting the
spectra with XSPEC it is opportune to allow the metallicities to
assume negative values. This procedure is necessary to avoid
underestimates on the derived metal abundances which could affect the
measurements especially for the case of low metallicity, statistically
poor spectra (for a more detailed discussion of this point see
Appendix A in \citealt{leccardi08_ab}).

Abundances are measured relative to the solar photospheric values of
\citet{anders89}, where Fe $= 4.68\times 10^{-5}$, Si $=
3.55\times 10^{-5}$ and Ni $= 1.78\times 10^{-6}$ (by number relative
to H). We have chosen these values to allow direct comparison with
other works present in the literature.  Spectra from all three EPIC
instruments (MOS1, MOS2 and {\it pn}) were fit individually.

In the course of our analysis we have found that spectra with less
than about 3500 source counts could not be used to constrain the
abundances of nickel and silicon, we therefore have decided to
eliminate from the sample all clusters having less than 3500 source
counts in the adopted energy band (i.e. 1.8 -- 10 keV) in at least one
of the three EPIC instruments. The five clusters removed from the
sample by adopting this criterion are: Abell 1644, Abell 1651, Abell
2065, Abell 2063 and Abell 3562.

We have also excluded Cygnus A because of its peculiar core
emission. A powerful double-lobed radio galaxies, QSO B1957$+$405
featuring huge jets feeding very large hotspot regions which are well
detected in X-rays, resides in the core of this cluster.  In the light
of the difficulty of removing efficiently the emission associated to
the radio galaxy from the spectra of the core region we have preferred
to exclude this cluster from the sample.

In the case of the nickel abundance measurement we have introduced a
further selection criterion to exclude clusters which are background
dominated in the hard part of the spectrum. We have selected the $7-9$
keV energy band, redshifted at the clusters distance, around the
nickel line at $\sim 8$ keV, and we have measured the relative
difference between the source and background count rates,
$(cr_{sou}-cr_{bgk})/cr_{bkg}$, in that band. We have then considered
only nickel measurements from clusters with relative difference larger
than 1, namely clusters with source count rates which were at least
twice the background count rates in that hard band.  Applying this
selections Virgo, Abell 262 and Abell 1060 are also excluded from the
sample. 
It is worth noting that our {\it pn} spectra are only mildly
contaminated by the fluorescence line complex located around 8 keV
(e.g. \citealt{xmm_uhb09}), the reason being that our spectra are
extracted from the central regions of the {\it pn} detector where, due
to a hole in the {\it pn} electronics box, the fluorescence lines are
weak.

In summary we have measured Si and Fe abundances from a
sample of 26 clusters and Ni from a subsample of 23 systems.

\subsection{Addendum to the spectral analysis of the pn}

\citet{molendi09} analyzing the {\it pn} spectra of
a long {\it XMM-Newton} observation of the Perseus cluster first noted
in the best-fit residuals the presence of a substantial structure
around the Fe K$\alpha$ line (see Sect. 3.1.1 and Fig. 2 in their
paper). This structure is attributed to an incorrect modeling of the
{\it pn} spectral resolution within the redistribution matrix.

Investigating this features extensively in the {\it pn} spectra of our
cluster sample we have found that the resolution miss-calibration can
be compensated for by including a multiplicative component that
performs a gaussian smoothing of the spectral model ({\it gsmooth} in
XSPEC). We set the width of the gaussian kernel to be 4 eV (FWHM) at 6
keV and assume a power-law dependency of the width on the energy with
an index of ~1.

Figure 1 shows the {\it pn} spectra and best-fit residuals for three
bright clusters: Perseus, 2A 0335$+$096 and Centaurus clusters. We
have fitted the spectra with the 4T model without and with the
gaussian smoothing. In the former cases (left panels in the Figure)
prominent residuals are present in the ``wings'' of the Fe line, while
in the latter (right panels) the residuals are significantly reduced.
In 13 (11) out of 26 cases the modified models 
applied to {\it pn} spectra provide a substantial better fit (i.e.
$\Delta \chi^2 > 2.7 (4.0)$) than the un-modified models.  We note
that Si and Ni lines do not show similar residuals with respect to the
best fit model, most likely because the statistical quality of the
data is not as high (Si and Ni) and because the spectral resolution is
significantly poorer (Si).

After the application of the gaussian smoothing the cross-calibration
between the {\it pn} and MOS improve.  For instance, the relative
differences, computed over the whole sample, between iron abundances
estimated from {\it pn} and MOS1 spectra, i.e.  $({\rm
Fe_{MOS1}-Fe}_{pn})/{\rm Fe}_{pn}$, with the un-smoothed and smoothed
4T model decrease from $12\%$ to $7\%$, whereas the ones between {\it
pn} and MOS2 decrease from $9\%$ to $4\%$ (errors on the given
percentages are always $1\%$).
 
\subsection{Choice of the best spectral model for each cluster}

In this section we discuss how we have selected the model that best
fits the spectral data among the 1T, 2T and 4T models described above.

Applying the statistical F-test between 1T and 2T (or 1T and 4T)
models, we find that only the first 6 out of 26 clusters with larger
statistics (i.e. Perseus, Virgo, 2A $0335+96$, Centaurus, Abell 478
and Abell 4038) show overwhelming evidence of multi-temperature
structure, namely an F-test probability, $PF$, smaller than $1\%$ for
all EPIC instruments.

The majority of the other clusters (17/26 for 2T model and 22/26 for
4T model) however, show a significant improvement ($PF<1\%$) in at
least one of the EPIC instruments.  For these clusters the relative
differences between the metal abundances measured in each instrument
with the 1T and 2T (or 1T and 4T) models are always smaller than
2-3$~\sigma$ for Fe and Si, and, consistent with zero for Ni.
Therefore, from a purely statistical point of view, the choice of 1T
or multi-temperature (2T or 4T) models results in modest differences.
Nevertheless, we point out that our spectra are extracted from the
very central regions of cool-core clusters and that these regions
always display temperature gradients (e.g. \citealt{leccardi08_t} and
references therein). It follows that the best description of the
spectrum of this plasma is through a multi-temperature model.  For
this reason and for the F-test results above, we decide to use a
multi-temperature modeling for all the clusters in the sample.

An F-test between 2T and 4T models is not possible as the two models
have the same numbers of degrees of freedom. However, relative
differences between the metal abundances measured with the two models
show that the systematic uncertainties associated to the different
modeling of the data are below $2-3\%$ for Fe and Si and consistent
with zero for Ni. These systematic errors are of the same order or
smaller than systematic errors on the abundances due to calibration
differences between the three EPIC detectors (systematic errors given
by the cross correlation between EPIC instruments will be discussed
hereinafter in Sect. 4.3).  Therefore the two models are substantially
equivalent to describe our data.  We choose to use for all the
clusters the 4T model.

 \begin{table}
 \begin{center}
 \caption{Measurements of silicon, iron and nickel abundances relative to
 the solar value \citep{anders89} for the final sample (26 clusters for
 Si and Fe, 23 for Ni) from the 4T {\it mekal} spectral model. Clusters
 are sorted by decrescent counts statistics in the {\it pn}. Errors on
 abundances include a $3\%$ systematic error as detailed in Sect. 4.3.}
 \label{tab_results}
 \begin{tabular}{r@{\hspace{.8em}}l@{\hspace{.8em}} c@{\hspace{.6em}} 
c@{\hspace{.8em}} c@{\hspace{.8em}} } \hline \hline
Seq. & Cluster & Si & Fe & Ni \\
\hline
1  & Perseus       &  $0.77\pm  0.03$ & $0.49\pm  0.02$ & $1.27\pm  0.11$ \\
2  & Virgo(M87)    &  $0.95\pm  0.07$ & $0.41\pm  0.02$ & $             $ \\
3  & 2A $0335+096$ &  $0.73\pm  0.04$ & $0.48\pm  0.02$ & $1.03\pm  0.06$ \\
4  & Centaurus     &  $1.62\pm  0.16$ & $0.94\pm  0.03$ & $2.50\pm  0.17$ \\
5  & Abell 478     &  $0.36\pm  0.08$ & $0.43\pm  0.02$ & $0.62\pm  0.16$ \\
6  & Abell 1795    &  $0.68\pm  0.04$ & $0.47\pm  0.02$ & $0.89\pm  0.11$ \\
7  & Abell 2597    &  $0.40\pm  0.09$ & $0.37\pm  0.02$ & $0.79\pm  0.51$ \\
8  & Abell 4038    &  $0.49\pm  0.08$ & $0.42\pm  0.03$ & $1.50\pm  0.23$ \\
9  & Abell 1060    &  $0.73\pm  0.07$ & $0.42\pm  0.02$ & $             $ \\
10 & MKW3s         &  $0.78\pm  0.07$ & $0.48\pm  0.02$ & $1.36\pm  0.75$ \\ 
11 & Abell 2052    &  $0.77\pm  0.03$ & $0.55\pm  0.02$ & $1.35\pm  0.09$ \\ 
12 & Abell 2199    &  $0.92\pm  0.10$ & $0.45\pm  0.02$ & $1.11\pm  0.41$ \\ 
13 & Abell 2029    &  $0.61\pm  0.15$ & $0.58\pm  0.03$ & $1.66\pm  0.86$ \\ 
14 & Abell 3112    &  $0.77\pm  0.10$ & $0.58\pm  0.04$ & $1.17\pm  0.43$ \\ 
15 & HYDRA A       &  $0.46\pm  0.11$ & $0.39\pm  0.02$ & $0.75\pm  0.13$ \\ 
16 & Abell 496     &  $0.85\pm  0.06$ & $0.55\pm  0.02$ & $1.72\pm  0.37$ \\ 
17 & AWM7          &  $1.13\pm  0.07$ & $0.64\pm  0.02$ & $1.24\pm  0.30$ \\ 
18 & ABELL 4059    &  $0.78\pm  0.03$ & $0.59\pm  0.04$ & $1.00\pm  0.12$ \\ 
19 & Abell 3571    &  $0.52\pm  0.15$ & $0.48\pm  0.03$ & $1.81\pm  0.09$ \\ 
20 & Abell 1650    &  $0.36\pm  0.16$ & $0.57\pm  0.03$ & $1.11\pm  0.62$ \\ 
21 & Abell 1689    &  $0.40\pm  1.68$ & $0.33\pm  0.03$ & $1.96\pm  0.33$ \\ 
22 & PKS $0745-191$&  $0.61\pm  0.22$ & $0.43\pm  0.03$ & $0.49\pm  0.22$ \\ 
23 & Abell 262     &  $0.93\pm  0.06$ & $0.53\pm  0.06$ & $             $ \\
24 & Abell 2204    &  $0.83\pm  0.24$ & $0.53\pm  0.03$ & $1.27\pm  0.49$ \\ 
25 & Abell 85      &  $0.94\pm  0.23$ & $0.55\pm  0.02$ & $2.49\pm  0.74$ \\ 
26 & Abell 3558    &  $0.84\pm  0.36$ & $0.49\pm  0.02$ & $1.68\pm  1.15$ \\ 
\hline
\end{tabular}

 \end{center}
 \end{table}

\subsection{EPIC Cross-calibration Issues}

For each cluster we compute the relative differences between iron,
silicon, and nickel measured with MOS1, MOS2 and {\it pn}, and
subsequently compute the weighted averages of these differences for
the whole sample (i.e. 26 clusters for Fe and Si, and 23 clusters for
Ni, results are from 4T model).  The resulting mean values for Si, Fe
and Ni are shown in Table \ref{tab_cross}.
From the results shown in this table it is clear that systematic
errors are often dominant with respect to statistical ones for Fe and
Si, whereas Ni appears to be fully dominated by statistical
uncertainties.  Another important point is that the relative
differences between MOS1 and MOS2 are smaller with respect to the
differences between MOS1 or MOS2 and {\it pn}.

To account for systematic differences between measures obtained from
different detectors and spectral models we sum in quadrature a 3\%
systematic error to each Fe and Si abundance measurement.  The
relative differences between iron, silicon, and nickel measured with
MOS1, MOS2 and {\it pn} computed over the whole sample, including the
3\% systematic errors, are smaller than those computed without them
and always significant at less than 3$\sigma$ (these values are given
in the second line relative to each element in Table \ref{tab_cross}).

For each cluster we compute an EPIC Fe and Si abundance by performing
error weighted averages over the three detectors, with errors
including the 3\% systematic described above.  Errors on the EPIC Fe
and Si abundances are computed by dividing the error weighted standard
deviation of the EPIC abundance by the square root of 3. To avoid
errors on the EPIC abundances from falling below the systematic level
a $3\%$ systematic error is summed in quadrature.

As shown in Table \ref{tab_cross} nickel measurements are clearly
dominated by statistical  errors, nevertheless we decided for internal
consistency to compute the nickel  value for each cluster as described
above  for Si  and Fe. Indeed  we expect  Ni to  be prone to the
same systematics affecting Si and Fe.

The distribution of Fe and Si abundances is not very symmetric, this
entails that it is not particularly well represented by a gaussian. We
have determined that the major cause is the presence of a few
measurements with extremely small errors.  Introducing a systematic
error of 3\% (see above) alleviates the problem considerably.  The
fact that the data is not normally distributed implies that quantities
such as the mean and standard deviation do not enjoy the properties
they "normally" do. A manifestation of non-normal behavior can be
observed in Table 2: consider for example the Fe abundance measured
with {\it mekal}, although the mean relative difference for MOS1 and
MOS2 measures is small 2\%, and the mean relative difference for MOS1
and {\it pn} is large, e.g. 7\%, the mean relative difference for MOS2
and {\it pn} is modest, e.g. 2\%. Similar results are observed for Si
measured with {\it mekal} and with Fe and Si measured with {\it apec}.

The final abundances are reported in Table \ref{tab_results}: column
(1) is the cluster name, column(2), (3) and (4) are the EPIC error
weighted averages of Si, Fe and Ni, respectively, with their $1\sigma$
 errors.

For completeness we have also computed all the abundances presented in
Table \ref{tab_results} using the 2T model, as expected we found only
modest differences with respect to those estimated with the 4T
model. The relative differences between the mean Fe, Si and Ni values
computed for the whole sample with the 2T and 4T models are $-3\%\pm
1\%$, $2\%\pm 7\%$ and $-1\%\pm 9\%$, respectively. In the case of Fe
we detect a systematic which is of the same order of the one
associated to the choice of detector, while in the case of Si and Ni
indeterminations are sufficiently large to hide systematics of the
same order.

  \begin{figure*} \centering
  \includegraphics[width=10cm,angle=-90]{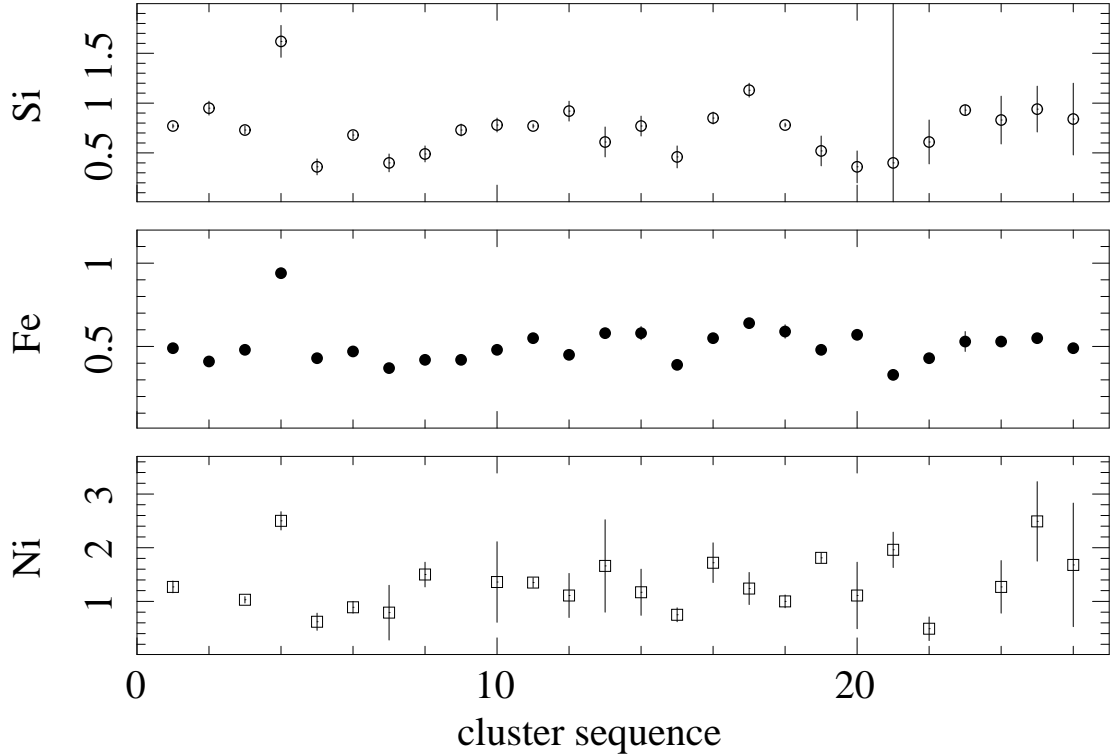}
  \caption{Silicon (upper panel), iron (medium panel) and nickel
  (lower panel) abundances measured in the core region with r $< 0.5$
  r$_{\rm cool}$ for the cluster sample. The total number of clusters
  is 26 for Si and Fe, and 23 for Ni. Clusters are sorted by
  decreasing {\it pn} counts statistics according to Table
  \ref{tab_results}. Weighted mean values and intrinsic scatters of
  the relations are given in Table \ref{tab_mean}.}
  \label{fig_sifeni} 
  \end{figure*}

\subsection{apec vs. mekal}

{\it Mekal} is not the only collisional ionization equilibrium (CIE)
emission model in XSPEC.  For quite some time an alternative plasma
code known as {\it apec} \citep{smith01_apec} has been available. Some
authors have performed comparisons between spectral fits performed
with the two codes \citep[e.g.][]{sanders06,deplaa07}.
\citet{sanders06}, analyzing a long {\it Chandra} observation
of the core of Centaurus find that while there are little differences
between Si and Fe abundances estimated with the two codes, the Ni
abundance found with {\it apec} is somewhat smaller than that found
with {\it mekal}, a similar result, albeit with much less statistics,
is found by de Plaa and collaborators. Here we wish to investigate how
the differences between the plasma codes impact on our abundance and
abundance ratio measurements. To this end, we reanalyze all our
objects with a multi-temperature model ({\it vapec+vapec+vapec+vapec})
analogous to the 4T {\it mekal} model which we dub 4T {\it apec}.  A
comparison amongst {\it apec} metal abundances measured with the three
EPIC instruments (see Tab. \ref{tab_cross}) provides results that are
very similar to those found for {\it mekal}; including a systematic of
3\% on the measures alleviates difference between the three
instruments in much the same way (see Tab. \ref{tab_cross}) it does
for the 4T {\it mekal} fits.  Mean metal abundances for Si, Fe and Ni
and abundance ratios, Si/Fe and Ni/Fe obtained with {\it apec} and
associated intrinsic scatters are reported in Tab. \ref{tab_mean}.
Comparing the sample averaged metal abundances measured with {\it
mekal} with those obtained with {\it apec} we find that: Fe is almost
unchanged,
Fe$_{apec}$/Fe$_{mekal} = 1.05\pm 0.01$; Si is somewhat higher,
Si$_{apec}$/Si$_{mekal} = 1.11\pm 0.02$ and Ni is lower,
Ni$_{apec}$/Ni$_{mekal} = 0.82\pm 0.04$.
The Si/Fe ratio, as measured with {\it apec}, is slightly larger than
that estimated with {\it mekal},
(Si/Fe)$_{apec}$/(Si/Fe)$_{mekal} = 1.06\pm 0.02$,
while Ni/Fe is substantially smaller,
(Ni/Fe)$_{apec}$/(Ni/Fe)$_{mekal} = 0.77\pm 0.04$.

 \begin{table}
 \begin{center}
 \caption{Averages of Si, Fe, Ni abundances and Si/Fe, Ni/Fe ratios
 for the cool-core cluster sample in solar units. Mean values are given
 both for the total sample and with the exclusion of the peculiar Centaurus
 cluster. Average abundance values and intrinsic scatters are computed
 with the maximum likelihood method described in \citet{maccacaro88}.}
 \label{tab_mean}                   
 \begin{tabular}{l c c}              
\hline\hline                        
{\it mekal} model & & \\
\hline
Metal & mean & scatter  \\
\hline
    Si          & $0.75\pm 0.03$ & $0.22\pm 0.03$ \\
    Si no Cent. & $0.72\pm 0.03$ & $0.17\pm 0.02$ \\
\hline
    Fe          & $0.51\pm 0.02$ & $0.11\pm 0.01$ \\
    Fe no Cent. & $0.49\pm 0.01$ & $0.07\pm 0.01$ \\
\hline
    Ni          & $1.28\pm 0.08$ & $0.45\pm 0.06$ \\
    Ni no Cent. & $1.18\pm 0.07$ & $0.34\pm 0.06$ \\
\hline
 Si/Fe          & $1.47\pm 0.05$ & $0.27\pm 0.05$ \\
 Si/Fe no Cent. & $1.45\pm 0.05$ & $0.30\pm 0.05$ \\
\hline
 Ni/Fe          & $2.41\pm 0.13$ & $0.60\pm 0.12$ \\
 Ni/Fe no Cent. & $2.40\pm 0.14$ & $0.63\pm 0.13$ \\
\hline
& & \\
& & \\
\hline\hline                        
{\it apec} model & & \\
\hline
Metal & mean & scatter  \\
\hline
    Si          & $0.83\pm 0.04$ & $0.23\pm 0.03$ \\
    Si no Cent. & $0.80\pm 0.03$ & $0.19\pm 0.03$ \\
\hline
    Fe          & $0.53\pm 0.02$ & $0.12\pm 0.01$ \\
    Fe no Cent. & $0.51\pm 0.01$ & $0.08\pm 0.01$ \\
\hline
    Ni          & $1.07\pm 0.10$ & $0.53\pm 0.07$ \\
    Ni no Cent. & $0.96\pm 0.08$ & $0.39\pm 0.06$ \\
\hline
 Si/Fe          & $1.56\pm 0.05$ & $0.25\pm 0.05$ \\
 Si/Fe no Cent. & $1.56\pm 0.05$ & $0.26\pm 0.05$ \\
\hline
 Ni/Fe          & $1.88\pm 0.14$ & $0.67\pm 0.14$ \\
 Ni/Fe no Cent. & $1.83\pm 0.15$ & $0.67\pm 0.15$ \\
\hline
\end{tabular}

 \end{center}
 \end{table}

In summary, while for Fe and Si/Fe the systematics associated to the
cross-calibration between EPIC experiments, 3\%, are comparable to
those related to the emission model, 5\% and 6\% respectively, for Si,
Ni, and Ni/Fe the dominant source of error are the systematics related
to the emission model, 11\%, 18\% and 23\% respectively. It goes
almost without saying that the statistical errors on the mean Si, Fe,
Ni, Si/Fe and Ni/Fe are always dominated by systematics.

\section{Abundances results in Cool-Cores}

\subsection{Mean Iron, Silicon and Nickel Abundances}

In Fig. \ref{fig_sifeni} we show the abundances of silicon (upper
panel), iron (middle panel) and nickel (lower panel) in the cool-cores
of our clusters (we use hereinafter results from the {\it mekal} model
although results from both spectral codes are reported in
Tab. \ref{tab_mean}).  Visual inspection of the plots indicates that
the abundance distributions show only small, 20\%-30\% variations with
respect to the mean.  The only object that is clearly out of the
distributions is the Centaurus cluster whose core is uncommonly rich
in metals \citep[e.g.][]{sanders06}.

  \begin{figure*} \centering
  \includegraphics[width=10cm,angle=-90]{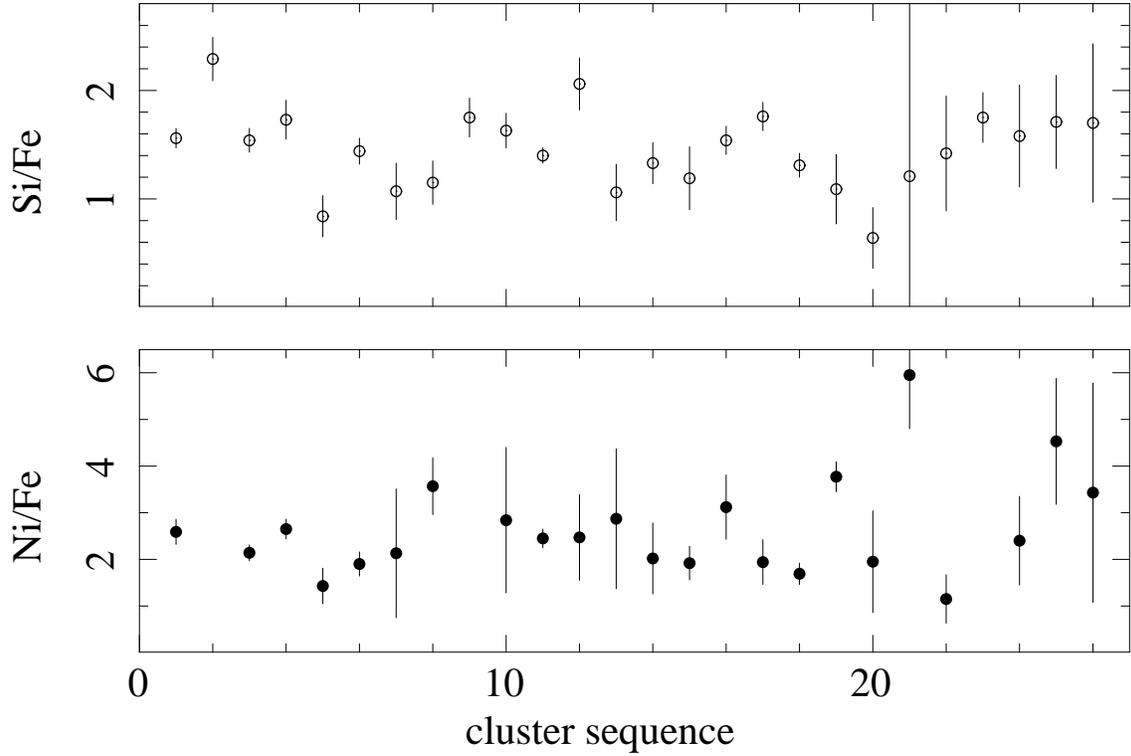}
  \caption{Silicon-to-iron (upper panel) and nickel-to-iron (lower
  panel) ratios for the cluster sample.  }
  \label{fig_ratio}
  \end{figure*}

To quantify the qualitative indications shown in Fig. \ref{fig_sifeni}
we have estimated the properties of the abundance distributions with a
maximum likelihood algorithm that postulates a gaussian parent
distribution described by a mean and an intrinsic dispersion
\citep{maccacaro88}.  Our analysis shows that our abundance
distributions feature a substantial intrinsic dispersion, roughly 30\%
for Si and Ni and 20\% for Fe. Best fitting means and intrinsic
dispersions for all species are summarized in Table \ref{tab_mean}.
To examine the influence of the outlier Centaurus cluster on the
averages and scatters, we have redone the computation excluding it
from the sample, results are reported in Tab. \ref{tab_mean}. We do
not find any dramatic change in the derived quantities, especially in
the scatters of the abundance distribution that are only slightly
smaller than those computed with the Centaurus cluster.

We have compared our Fe abundance estimates with those from
\citet{leccardi09_ab} who made measures in a circular region with
radius 0.05 r$_{180}$, which roughly corresponds to our 1/2 r$_{\rm
cool}$ extraction radius.  We have considered their cool-cores systems
excluding objects contained in our sample (these selections leads to a
subsample of 13 clusters located in the 0.1-0.25 redshift
interval). Applying the maximum likelihood algorithm to their Fe
distribution we find a mean Fe abundance of $0.49\pm 0.02$ and a
relative intrinsic scatter of $16\% \pm 3\%$, which are both in
excellent agreement with our measurements.

The mean Si and Fe abundance values reported in Table
\ref{tab_mean} are also in good agreement with what found by
\citet{tamura04} for the innermost regions  of their cool-core
clusters with temperatures larger than 3 keV (see Table 6 in
\citealt{tamura04}).

On the contrary \citet{deplaa07} found Si, Fe and Ni abundances
somewhat smaller than ours.  By applying the maximum likelihood method
to the abundances given in their Table A.1 \citep[see][]{deplaa07}, we
find an average Si of $0.39\pm 0.02$, Fe of $0.35\pm 0.02$ and Ni of
$0.76\pm 0.05$ \citep[here and in the rest of this work abundances
taken from the literature are rescaled to][]{anders89}, with intrinsic
scatters for Si and Fe of 24\% and 17\%, respectively. The smaller
abundance values are not surprising since the extraction regions used
by \citet{deplaa07} are about three times larger than those considered
in our work and because their sample contains both cool-core and non
cool-core clusters.

Interestingly, the mean Si and Fe found by \citet{rasmussen07} from a
sample of galaxy groups are very similar to ours. From their Fig. 11
(panels b and c) we estimate a mean Si of $\sim 0.75$ and a mean Fe
of $\sim 0.50$ within 0.08 r/r$_{500}$ which roughly corresponds to
our 1/2 r$_{\rm cool}$ extraction radius.

\subsection{Silicon-to-Iron and Nickel-to-Iron ratios}

In Fig. \ref{fig_ratio} we plot the silicon-to-iron (upper panel) and
nickel-to-iron (lower panel) ratios distributions for our
objects. Using the maximum likelihood algorithm described in the
previous section we have computed the mean and the intrinsic
dispersion for the Si/Fe and Ni/Fe distributions, results are shown in
Tab. \ref{tab_mean}.  The relative scatter found for Si/Fe is about
20\%, i.e. comparable to the one estimated for Fe and somewhat smaller
than the one for Si; for Ni/Fe the scatter is only slightly larger
25\%.  Interestingly the Centaurus cluster, which features very high
central abundances, is characterized by Si/Fe and Ni/Fe ratios very
similar to those observed in other clusters.

As expected from the results on the Si and Fe abundances shown in the
previous Section we find that our average Si/Fe ratio is in good
agreement with the value found by \citet{tamura04}, $0.16\pm 0.2$, in
the innermost bin of their high temperature ($>3$ keV) cool-core
clusters sample, whereas in \citet{deplaa07} the mean Si/Fe and Ni/Fe
abundance ratios are $1.07\pm 0.03$ and $2.07\pm 0.13$, which are
significantly lower than our values (the abundance ratios are computed
with the maximum likelihood method applied to the data given in Tab
A.1 of \citealt{deplaa07}). As already mentioned, the observed
difference between our abundance ratios and those from
\citet{deplaa07} could derive from the different sample selection.

We have compared our Si/Fe ratio with those measured for a sample of
groups \citep{rasmussen07} and of X-ray luminous elliptical galaxies
\citep{humphrey06}. We note that in both papers abundances are
reported in solar units which differ from ours, \cite{grevesse98} for
\cite{rasmussen07} and \cite{asplund05} for \cite{humphrey06};
to allow immediate comparison with our results the values shown below
have been converted to \cite{anders89} units. The measure on the
groups at scales comparable to those adopted for our clusters, 0.1
r$_{500}$, (see Fig. 11 f and Sect. 5.3 in \citealt{rasmussen07}),
provides a mean Si/Fe ratio of 1.35 and a relative scatter of $32\%$
which are both in broad agreement with ours.  The measure on the
galaxies was made using data from high-luminosity elliptical reported
in Table 2 by \cite{humphrey06}. We have applied the maximum
likelihood method used for our own data deriving a mean Si/Fe value of
$1.50\pm 0.07$ and a relative intrinsic scatter of $16\% \pm 3\%$,
which are both in good agreement with our estimates for cool-cores.

Summarizing, the average Si/Fe ratio appears to be nearly constant
from the galactic to the rich clusters scale suggesting a common
enrichment scenario in these objects.

\section{SNIa vs. SNcc}

There are indications that both type Ia and core-collapsed supernovae
contribute significantly to the enrichment of the intergalactic medium
in cores of cool-core clusters
\citep[e.g.][]{ishimaru97,finoguenov00,gastaldello02}.  We try here to
determine the relative proportion between the two SNe types using the
silicon-to-iron and nickel-to iron abundance ratios normalized to the
solar value.

The observed X$_i$/Fe ratio, where X$_i$ is the $i$-th element, can be
expressed as a linear combination of the X$_i$/Fe ratio expected from
type Ia and core-collapsed supernovae \citep[e.g.][]{gastaldello02}:

$$ \biggl ( \frac{{\rm X}_i}{\rm Fe} \biggr )_{\rm obs} = \xi \cdot
\biggl ( \frac{{\rm X}_i}{\rm Fe} \biggr )_{\rm SNIa} + (1 - \xi) \cdot
\biggl ( \frac{{\rm X}_i}{\rm Fe} \biggr )_{\rm SNcc}, \eqno(1) $$

where
$$\xi = \frac{M_{\rm Fe,SNIa}}{M_{\rm Fe,SNIa} + M_{\rm Fe,SNcc}} \eqno(2) $$

is the SNIa iron-mass-fraction, i.e. the fraction of the iron mass
synthesized in type Ia supernovae, $M_{\rm Fe,SNIa}$ and $M_{\rm
Fe,SNcc}$ are respectively the total Fe mass ejected by SNIa and SNcc.

 \begin{table*}
 \begin{center}
 \caption{Supernovae type Ia mass fraction, $\xi =
 M_{Fe,SNIa}/(M_{Fe,SNIa}+M_{Fe,SNcc})$, obtained for different
 theoretical SNe silicon-to-iron and nickel-to-iron ratios from the
 observed Si/Fe and Ni/Fe ratios. SNIa yields are computed from the
 full set of models for deflagration (W7, W70) and delayed detonation
 (WDDs, CDDs) scenarios by \citet{iwamoto99}. SNcc yields are
 integrated over a Salpeter ($x=1.35$) or a AY top-heavy ($x=0.95$)
 IMF, with abundance of the progenitor Z=0.02. Models for SNcc are
 taken from \citet[][No06]{nomoto06}, \citet[][CL04]{chieffi04} and
 \citet[][W95]{woosley95}.}
 \label{tab_xi}               
 \begin{tabular}{r c c c c c c c}   
\hline\hline                       
{\it mekal} model ~~~~~~\vline~~~~& & & & SNIa &  & &\\
\hline 
SNcc   ~~~~~~\vline~~~~& W7           & W70          & WDD1         & WDD2         & WDD3         & CDD1         & CDD2 \\
\hline 
No06,S ~~~~~~\vline~~~~&$0.528- 0.571$&$0.515- 0.734$&$            $&$            $&$            $&$            $&$            $ \\
No06,T ~~~~~~\vline~~~~&$0.575- 0.579$&$0.562- 0.760$&$            $&$            $&$            $&$            $&$            $ \\
CL04,S ~~~~~~\vline~~~~&$            $&$0.505- 0.582$&$            $&$0.546- 0.789$&$0.505- 0.728$&$            $&$0.534- 0.770$ \\
CL04,T ~~~~~~\vline~~~~&$            $&$0.559- 0.728$&$            $&$0.600- 0.727$&$0.559- 0.758$&$            $&$0.588- 0.727$ \\
W95,S  ~~~~~~\vline~~~~&$            $&$            $&$0.616- 0.677$&$0.518- 0.775$&$0.476- 0.712$&$0.635- 0.652$&$0.505- 0.756$ \\
W95,T  ~~~~~~\vline~~~~&$            $&$            $&$            $&$0.554- 0.792$&$0.512- 0.732$&$            $&$0.541- 0.774$ \\
\hline 
& & & & & & & \\
& & & & & & & \\
\hline\hline                    
{\it apec} model ~~~~~~\vline~~~~& & & & SNIa &  & &\\
\hline 
SNcc   ~~~~~~\vline~~~~& W7          & W70         & WDD1         & WDD2         & WDD3         & CDD1         & CDD2 \\
\hline 
No06,S ~~~~~~\vline~~~~&$           $&$0.475-0.511$&$            $&$0.514- 0.764$&$0.475- 0.707$&$            $&$0.502- 0.747$\\
No06,T ~~~~~~\vline~~~~&$           $&$           $&$            $&$0.564- 0.789$&$0.526- 0.735$&$            $&$0.553- 0.773$\\
CL04,S ~~~~~~\vline~~~~&$           $&$           $&$0.591- 0.886$&$0.502- 0.759$&$0.581- 0.701$&$0.608- 0.841$&$0.490- 0.741$\\
CL04,T ~~~~~~\vline~~~~&$           $&$           $&$0.647- 0.855$&$0.561- 0.787$&$0.523- 0.734$&$0.663- 0.799$&$0.549- 0.771$\\
W95,S  ~~~~~~\vline~~~~&$           $&$           $&$0.560- 0.885$&$0.640- 0.744$&$            $&$0.577- 0.885$&$0.640- 0.725$\\
W95,T  ~~~~~~\vline~~~~&$           $&$           $&$0.598- 0.895$&$0.561- 0.763$&$            $&$0.615- 0.868$&$0.561- 0.745$\\
\hline
\end{tabular}

 \end{center}
 \end{table*}

\cite{matteucci05} caution against an inappropriate use of equation (1)
as in principle this equation is valid only if one is interested in
the global Fe production (i.e., Fe in stars, galaxies and the ICM).
Indeed the model leading to eq. (1) does not consider the finite
timescale over which the gas is processed through stars and ignores
the mechanism of chemical enrichment of the ICM from
galaxies. However, as already noted by \cite{deplaa07} and
\cite{humphrey06}, eq. (1) can be applied to the ICM, quite simply the
iron mass fractions that are derived from it should be interpreted as
the iron mass fractions that would be needed to enrich the ICM, not as
the actual iron mass fraction produced throughout the history of the
cluster.

By inserting Si/Fe and Ni/Fe ratios expected for SNIa and SNcc and the
mean observed Si/Fe and Ni/Fe ratios in eq. (1) we estimate the SNIa
iron-mass-fraction $\xi$.  We compute $\xi$ by combining several
theoretical SNe yields, the ones for SNIa originate from two
physically different sets of models taken from \citet{iwamoto99},
namely the slow deflagration (W7 and W70) and the delayed detonation
(WDDs and CDDs) explosion models.  For core-collapsed supernovae we
use the yields from \citet{nomoto06}, \citet{chieffi04} and
\citet{woosley95}, details on the models and the computation of the
yields are provided in App. A.  Since we have two equations, one for
Si/Fe the other for Ni/Fe, with one unknown, $\xi$, we are in a
position to determine whether a given combination of Ia and
core-collapsed models can adequately reproduce the observed
ratios. One way of going about this \citep[e.g.][]{deplaa07} is to
perform a $\chi^2$ fit using the observed Si/Fe and Ni/Fe ratios and
associated errors as the data, the Si and Ni yields relative to Fe for
SNIa and SNcc, namely ${\rm (Si/Fe)_{SNIa}}$ and ${\rm (Si/Fe)_{SNcc}}$,
as constants and $\xi$ as fitting parameter.
We prefer to proceed differently, indeed it has been noted that the
uncertainties associated to the SNe yields are in the order of tens of
percent (e.g., \citealp{woosley95,gibson97}, P. Young priv. comm., and
the scatter in the yields we have gathered from the literature fully
confirms this, see Table \ref{tab_yields_SNe} and Appendix A), while
our observed mean values of Si/Fe and Ni/Fe have errors smaller than
5\%, if we, for the time being, neglect the uncertainties associated
to the spectral model. Consequently a robust estimate of $\xi$ should
first of all take into account the uncertainties in the yields. Given
the lack of precise information we have assumed a uniform
indetermination in the Si and Ni yields relative to Fe for SNIa and
SNcc of 20\%.  For each combination of SNIa and SNcc models we allow
the Si and Ni yields relative to Fe for SNIa and SNcc to vary by 20\%
and, using the observed Si/Fe and Ni/Fe ratios, we determine the range
of values of $\xi$ for which eq. (1) is satisfied. For some
combinations of SNIa and SNcc models, despite the generous 20\% range,
there are no values of $\xi$ satisfying eq. (1), which implies that
the observed Si/Fe and Ni/Fe cannot be reproduced for that particular
combination of SNIa and SNcc models.

In Table \ref{tab_xi} we report ranges of values for $\xi$ for all
possible combinations of Ia and core-collapsed supernovae models.  The
overall permitted range for $\xi$ is large $0.48 - 0.79$ and sensitive
to the indetermination in the SNe yield ratios, indeed assuming 10\%
or 30\% rather than 20\% we derive a range of $0.55 - 0.73$ and $0.37
- 0.85$, respectively.  Since the indetermination in the abundance
ratios related to the emission model is comparable to the one in the
yields we recompute the SNIa iron-mass-fraction assuming abundance
ratios derived from {\it apec} rather than {\it mekal}.  Results are
rather similar, see Table \ref{tab_yields_SNe}, the overall permitted
range for $\xi$ is similar $0.49 - 0.90$, the major difference is in
which of the combination of SNIa and SNcc models provide valid
solutions.

Equations (1) and (2) may be generalized to any atomic species and can
therefore be used to obtain the X$_i$ elements SNIa gas-mass-fraction.
We have done so and derived the Si and Ni SNIa gas-mass-fraction ranges
which are respectively $0.14 - 0.49$ and $0.26 - 0.88$.

 \begin{table*}
 \begin{center}
 \caption{Supernovae type Ia fraction, $f = SNIa/(SNIa+SNcc)$,
 obtained for different theoretical SNe models from the observed Si/Fe
 and Ni/Fe abundance ratios. SNIa and SNcc models are the same as in
 Table \ref{tab_xi}.}
 \label{tab_f}
 \begin{tabular}{r c c c c c c c}
\hline\hline                    
{\it mekal} model ~~~~~~\vline~~~~& & & & SNIa &  & &\\
\hline 
SNcc   ~~~~~~\vline~~~~& W7           & W70          & WDD1         & WDD2         & WDD3         & CDD1         & CDD2 \\
\hline 
No06,S ~~~~~~\vline~~~~&$0.117- 0.136$&$0.110- 0.244$&$            $&$            $&$            $&$            $&$            $\\
No06,T ~~~~~~\vline~~~~&$0.139- 0.141$&$0.132- 0.272$&$            $&$            $&$            $&$            $&$            $\\
CL04,S ~~~~~~\vline~~~~&$            $&$0.149- 0.193$&$            $&$0.168- 0.384$&$0.134- 0.289$&$            $&$0.154- 0.348$\\
CL04,T ~~~~~~\vline~~~~&$            $&$0.182- 0.320$&$            $&$0.204- 0.313$&$0.165- 0.327$&$            $&$0.188- 0.302$\\
W95,S  ~~~~~~\vline~~~~&$            $&$            $&$0.207- 0.254$&$0.129- 0.322$&$0.102- 0.237$&$0.226- 0.239$&$0.118- 0.289$\\
W95,T  ~~~~~~\vline~~~~&$            $&$            $&$            $&$0.143- 0.338$&$0.113- 0.250$&$            $&$0.131- 0.304$\\
\hline 
& & & & & & & \\
& & & & & & & \\
\hline\hline                    
{\it apec} model ~~~~~~\vline~~~~& & & & SNIa &  & &\\
\hline 
SNcc   ~~~~~~\vline~~~~& W7          & W70          & WDD1         & WDD2         & WDD3         & CDD1         & CDD2 \\
\hline
No06,S ~~~~~~\vline~~~~&$           $&$0.096- 0.109$&$            $&$0.107- 0.270$&$0.086- 0.200$&$            $&$0.098- 0.243$\\
No06,T ~~~~~~\vline~~~~&$           $&$            $&$            $&$0.130- 0.301$&$0.104- 0.225$&$            $&$0.119- 0.272$\\
CL04,S ~~~~~~\vline~~~~&$           $&$            $&$0.221- 0.606$&$0.144- 0.345$&$0.174- 0.262$&$0.239- 0.517$&$0.133- 0.313$\\
CL04,T ~~~~~~\vline~~~~&$           $&$            $&$0.269- 0.543$&$0.179- 0.388$&$0.145- 0.300$&$0.290- 0.453$&$0.166- 0.354$\\
W95,S  ~~~~~~\vline~~~~&$           $&$            $&$0.172- 0.556$&$0.197- 0.286$&$            $&$0.186- 0.564$&$0.189- 0.258$\\
W95,T  ~~~~~~\vline~~~~&$           $&$            $&$0.191- 0.574$&$0.147- 0.301$&$            $&$0.207- 0.517$&$0.141- 0.272$\\
\hline 
\end{tabular}

 \end{center}
 \end{table*}

By adopting abundance ratios derived from {\it apec} we find broader
ranges both for Si, $0.12 - 0.86$, and Ni, $0.01 - 0.90$.
The reason for this difference rests in the smaller Ni/Fe ratio found
with {\it apec}. More specifically the {\it mekal} value, 2.41, is
relatively high when compared to the Ni/Fe ratios predicted by SNIa
and SNcc models. This implies that the observed Ni/Fe and Si/Fe ratios
may only be reproduced by a limited combination of SNIa and SNcc
models. Conversely the {\it apec} value, 1.88, is closer to the mean
Ni/Fe ratio predicted by models and may be reproduced by a broader
combination of SNIa and SNcc yields.

The total Fe mass ejected by SNIa, $M_{\rm Fe,SNIa}$, and by SNcc,
$M_{\rm Fe,SNcc}$ can be rewritten
as:
$$M_{\rm Fe,SNIa} = N_{\rm SNIa}~ y_{\rm Fe,SNIa} \eqno(3)$$
and
$$M_{\rm Fe,SNcc} = N_{\rm SNcc}~ y_{\rm Fe,SNcc}, \eqno(4) $$

where $N_{\rm SNIa}$ and $N_{\rm SNcc}$ are respectively the total
number of type Ia and core-collapsed supernovae and $y_{\rm Fe,SNIa}$
and $y_{\rm Fe,SNcc}$ are respectively the Fe yields per SNIa and
SNcc.

We define, $f \equiv N_{\rm SNIa}/(N_{\rm SNIa}+N_{\rm SNcc})$ as the
number ratio of supernovae type Ia over the total number of
supernovae; combining eq. (2) with eqs. (3) and (4) we solve for $f$:

$$ f = \frac{\xi ~ y_{\rm Fe,SNcc}}{\xi ~ y_{\rm Fe,SNcc} + (1 - \xi)
~ y_{\rm Fe,SNIa}}. \eqno(5)$$

In Table \ref{tab_f} we report estimates for $f$ obtained by inserting
allowed ranges for $\xi$ in eq. (5).  We do not introduce any further
indetermination in the Fe yields as the scatter in Si and Ni yields is
much larger than the one in Fe and already accounted for through the
scatter introduced in the Si and Ni yields relative to Fe.  As for
$\xi$ the estimates are expressed in the form of permitted ranges. The
overall permitted range for $f$ is rather large: 0.10-0.38.  Using
abundance ratios estimated with the {\it apec} code we estimate a
somewhat broader permitted range of 0.09-0.61.

\section{Discussion}

In this work we have analyzed the central regions (r $<$ r$_{\rm
cool}/2$) of a sample of 26 nearby and rich cool-core clusters derived
from the original B55 cluster sample \citep{edge92}.  We have used
these regions characterized by the highest cluster photon statistics
to identify all the possible sources of systematic uncertainties in
the abundance measurements of the most relevant elements observable in
the X-ray spectral range between 1.8 and 10 keV, namely silicon, iron
and nickel.

Our analysis shows that, within the cool-cores of bright nearby
clusters, metal abundances of Si, Fe and in a few instances even Ni
can be measured to a high precision. Indeed the precision is so high,
particularly for Fe, that we need to introduce a cross-calibration
uncertainty of 3\% to reconcile measurements secured with different
EPIC experiments.  Thanks to the high statistical quality of our data
we find evidence for some intrinsic scatter in the element abundance
with Fe around 20\% and Si and Ni about 30\%, abundance ratios are
also characterized by roughly the same intrinsic scatter.

The relatively modest scatter around mean values of Si, Fe and Ni
abundances and Si/Fe and Ni/Fe ratios indicates that, whatever the
process responsible for the enrichment of the ICM, it works rather
similarly in all objects. Even in the somewhat extraordinary case of
the Centaurus cluster with its anomalously large metal content, the
abundances relative to iron are perfectly compatible with those of the
other clusters indicating the presence of similar enrichment
processes.  Uncertainties on the abundance estimates associated to the
specific choice of spectral model, i.e. 2T rather than 4T appear to be
negligible (i.e., within the statistical uncertainties for Si and Ni,
and of the same order of the EPIC cross-calibration differences for
iron).

The estimate of the abundances from our spectra may be thought of,
with some simplification, as a two step process: in the first step the
equivalent width of a given line is estimated; in the second step the
equivalent width is converted into an abundance assuming the
temperature derived by fitting the continuum. While the first step is
relatively straight-forward and, at least for K-shell lines, leaves
little room for ambiguity, the second, requiring estimates of
collisional ionization and transition probabilities is far more prone
to differences.  We have verified that the two plasma codes
available within XSPEC provide somewhat different estimates for the
metal abundances, the differences are not huge and qualitatively
similar to those found by other workers \citep{sanders06,deplaa07},
however, given the high quality of our abundance estimates, they
provide the dominant source of indetermination for Si (10\%), Ni
(16\%) and Ni/Fe (22\%) and contribute substantially in the case of Fe
(4\%) and Si/Fe (\%6).

We find that the Si/Fe abundance ratio estimated for cool-core clusters
\citep[this work;][]{tamura04}, shows a remarkably good agreement with the
values found from samples of galaxy groups \citep{rasmussen07} and
X-ray luminous elliptical galaxies \citep{humphrey06}. This similarity
favors a common enrichment scenario for luminous elliptical galaxies
and the cool-cores of groups and clusters, i.e. a similar mix of SNIa
and SNcc. 

We have used our estimates of the Si/Fe and Ni/Fe abundance ratios to
constrain the relative contribution of type Ia and core-collapsed
supernovae to the enrichment process. To this end we have compiled a
list of SNIa and SNcc yields from the literature. Simply by inspecting
our yields (see Tab. \ref{tab_yields_SNe} and \ref{tab_ratios_SNe})
it is rather obvious that the large differences, in the tens of \%,
are bound to have a non-negligible impact on our estimates. The Fe
normalized yields of Si and Ni both feature a scatter that is larger
than the indetermination in the corresponding observed abundance
ratios.  Under the rather simplistic assumption of a 20\%
indetermination in the yields we find that, $f$, the fraction of Ia to
Ia plus core-collapsed supernovae cannot be reconciled with 0 or 1, in
other words we need both Ia and core-collapsed supernovae to produce
the observed ratios. This result appears to be rather solid, an
indetermination of at least 50\% is required to allow $f=0$ and than
only for one specific combination of Ia and core-collapsed supernovae,
an indetermination of more than 70\% to have $f=1$.

Going beyond the qualitative statement that both SNIa and SNcc
contribute to the enrichment of the ICM is quite hard.  The accepted
range for $f$, the ratio of type Ia SN to all SNe, convolved over all
the SNIa and SNcc combinations providing valid results, is rather
large, from 10\% to 40\%.  Similarly we can say that: roughly more
than half (48-79\%) of the Fe is produced by SNIa; less than half of
the Si (14-49\%) is produced by SNIa and Ni is very poorly constrained
(26-88\%).  Increasing the indetermination on the yields beyond 20\%
will of course further enlarge the allowed ranges and result in even
weaker constraints.  We reiterate that the limiting factor is not the
X-ray abundance estimates, mean values can be constrained rather well,
even when allowing for systematics associated to instrument
calibration and thermal emission model uncertainties, but the large
indetermination in the theoretical yields. A similar point was made
some ten years ago by \citet{gibson97}, a decade later the
substantially improved observational constraints stand out in stark
contrast with the lack of any similar advancement on the theoretical
side.

Our analysis shows how difficult it is to determine the relative
contribution of type Ia and core-collapsed supernovae to the
enrichment of the ICM from global cool-core measurements.  An
alternative approach is to measure radial profiles of metal
abundances.  In this case the variation of an abundance ratio such as
Si/Fe, where the two elements are produced in different proportions by
SNcc and SNIa, can be interpreted as evidence for a variation of the
contribution of one type of SN with respect to the other.  This kind
of analysis has been attempted in the past,
\cite{finoguenov00,finoguenov01} using ASCA data, found that for most
rich clusters in his sample the Si/Fe rapidly increased when moving
from small to large radii.
Later work using {\it XMM-Newton} data did not confirm these findings,
\citet{tamura04} find a flat Si/Fe ratio for a sample of 19
clusters.  Amongst the few objects in common between the ASCA and
{\it XMM-Newton} samples are A3112 \citet{finoguenov00} and A2052
\citet{finoguenov01} for which ASCA data found Si/Fe radial
gradients and {\it XMM-Newton} derived flat Si/Fe profiles.
Recent analysis of {\it SUZAKU} data
\citep{sato07_a1060,sato09_a262,sato09_ngc507,komiyama09_ngc5044} on a
handful of groups and poor clusters leads to Si/Fe profiles that are
consistent with being constant out to at least 0.1r$_{180}$ and in
some instances to 0.2r$_{180}$.  Since at 0.1r$_{180}$ the average Fe
abundance excess is about 1/2 of what it is within 0.03r$_{180}$
\citep{leccardi08_ab}, this seems to rule out the possibility that the
transition from central excess to flat Fe profile might be associated
to a change in SN type mix.

Future work on {\it XMM-Newton}, {\it Chandra} and {\it SUZAKU} data
may change somewhat the situation by providing new abundance ratio
measurements beyond 0.1r$_{180}$. However, given the considerable
difficulties involved in estimating abundances at large radii
\citep[e.g.][]{leccardi08_ab}, it is by no means clear if and by how
much our estimates will improve.  A major advancement will be provided
by the first mission carrying a micro-calorimeter, most likely the
Japanese ASTRO-H. The ten fold increase in resolution at the Si line
will allow to reduce the background by an order of magnitude for the
H-like Si line and by a factor of a few for the He-like Si
triplet. This will translate into measures of Si abundances out to at
least 0.3r$_{180}$ for a good number of nearby clusters.

\section{Summary}

We have performed a detailed study of the Si, Fe and Ni abundances in
the cores of 26 nearby cool-core clusters. Our work may be divided
into two main parts, the first relates to the measure of the
abundances and the second to the estimate of the relative contribution
of SNIa and SNcc to the ICM enrichment process.
Regarding the first part our main result may be summarized as follows.

\begin {itemize}

\item
We find that systematic uncertainties associated to the different
spectral modeling, namely 2T versus 4T model, are below a few per cent
($2-3\%$).

\item
We find evidence for a $3\%$ systematic error between Si and Fe
abundance measures secured with the three EPIC detectors. The Ni
abundances do not have the statistical quality to investigate such
small systematic errors.

\item
We have verified that the {\it mekal} and {\it apec} plasma codes
available in XSPEC give somewhat different abundances values. Given
the high photon statistics of our spectra they contribute
significantly to the indetermination of Fe ($4\%$) and Si/Fe ($6\%$)
and are the dominant source of indetermination from Si ($\sim 10\%$),
Ni ($\sim 15\%$) and Ni/Fe ($\sim 20\%$).

\item
The final Si, Fe and Ni abundance distributions as well as the Si/Fe
and Ni/Fe abundance ratios distributions of the sample show only
moderate spreads (from $20\%$ to $30\%$) around their mean values
(exact values are reported in Tab. \ref{tab_mean}). These nearly
constant distributions suggest similar ICM enrichment processes at
work in all cluster cores.

\item
We find a remarkably uniformity in the observed Si/Fe abundance ratio
ranging from X-ray luminous elliptical galaxies to the cool-cores of
groups and clusters.  This tell us that, whatever the real proportion
between different SNe types may be, the enrichment process of the hot
gas associated to elliptical galaxies is likely the same in isolated
ellipticals, dominant galaxies in groups and brightest cluster
galaxies in clusters.

\end {itemize}

Regarding the second part, we have used our estimates of the Si/Fe and
Ni/Fe abundance ratios to constrain the relative contribution of SNIa
and SNcc to the enrichment process of the ICM. We have considered a
suite of 6 SNcc yields and 7 SNIa yields and, unlike previous studies,
we have included both uncertainties associated to the observed
abundance ratios {\it and} the theoretical yields for Si, Fe and Ni.
Under the assumption that the indetermination on the yields currently
available in the literature is of the order of $ 20\%$, we find that:

\begin {itemize}

\item
the SNIa iron-mass-fraction, $\xi$, overall permitted range is $0.48 -
0.79$;

\item
the SNIa silicon-mass-fraction and nickel-mass-fraction ranges are
$0.14 - 0.49$ and $0.26 - 0.88$, respectively;

\item
the number ratio of SNIa over the total number of SNe, $f$, spans from
0.10 to 0.38;

\item
the dominant source of uncertainty in the estimate of $\xi$ are errors
on theoretical yields, conversely, the choice of spectral code has
almost no impact.

\end {itemize}

Our analysis shows that the large uncertainties on the currently
available yield tables prevent any precise estimate of the relative
contribution of SNIa and SNcc, and that all that can really be said
with some certainty is that they both concur to the ICM enrichment
process in the cool-core regions.
An alternative approach to estimate the relative contribution of SNIa
and SNcc is to measure the difference between abundance ratios, such
as Si/Fe, in the core and in the outer regions of individual
systems. While early attempts have not provided definitive results, a
more robust analysis of currently available data may furnish important
constraints. The coming into operation of the first space-borne
micro-calorimeter, most likely ASTRO-H, will undeniably allow a
significant step forward.

\begin{acknowledgements}
We acknowledge useful discussions with Mariachiara Rossetti and Fabio
Gastaldello. This research has made use of the NASA/IPAC Extragalactic
Database (NED) which is operated by the Jet Propulsion Laboratory,
California Institute of Technology, of the NASA's High Energy
Astrophysics Science Archive Research Center (HEASARC), and, of the
{\it XMM-Newton} archive.
\end{acknowledgements}

\bibliographystyle{aa} 
\bibliography{biblio}  

\begin{appendix}

\section{}

We have computed yields for Si, Fe and Ni for different models of Ia
and core-collapsed supernovae.

The ones for SNIa originate from two physically distinct sets of
models taken from \citet{iwamoto99}, namely the slow deflagration (W7
and W70) and the delayed detonation (WDDs and CDDs) explosion
models. We have computed silicon, iron and nickel yields for a given
SNIa model by summing up over all stable isotopes the synthesized
amount of Si, Fe and Ni per SN event given in Table
\ref{tab_yields_SNe} of \citet{iwamoto99}.

For core-collapsed supernovae we use the yields from \citet{nomoto06},
\citet{chieffi04} and \citet{woosley95}.
In these cases we have first averaged the yields over the Initial Mass
Function (IMF) of the stellar population as follows:

$$ M_i = {{\int_{10 M_{\odot}}^{50 M_{\odot}} M_i(m)~m^{-(1+x)} dm}\over
{\int_{10 M_{\odot}}^{50 M_{\odot}} m^{-(1+x)} dm}}, \eqno(A1)$$

where $M_i(m)$ is the $i$-th element mass produced by a star of mass
$m$. We consider two IMF: a standard \citet{salpeter55} function with
exponent $x=1.35$, and, a top-heavy function with $x=0.95$
\citep[][thereafter AY]{arimoto87}, which
predicts a relatively larger number of massive stars. The integration
mass range is between 10 and 50 solar masses in agreement with
\citet{nomoto06}.

 \begin{table*}
 \begin{center}
 \caption{Yields of elements in solar masses for different SNe
 models. Type Ia yields were calculated for each element for the W7,
 W70, WDD1, WDD2, WDD3 and CDD1, CDD2 models in \citet{iwamoto99}. SN
 core-collapsed yields (including types Ib and Ic) were calculated from
 the results in \citet{nomoto06} integrating over a progenitor initial
 mass function. We have considered two different initial mass
 functions, the Salpeter function with index 1.35 and the AY top-heavy
 function with index 0.95, and we have integrated between 10 and 50
 solar masses. The initial metallicity of the progenitors is Z=0.02
 (i.e. solar).}
 \label{tab_yields_SNe}
 \begin{tabular}{  l@{\hspace{.8em}}  c@{\hspace{.6em}} 
c@{\hspace{.8em}} c@{\hspace{.8em}}  c@{\hspace{.8em}}  
c@{\hspace{.8em}} c@{\hspace{.8em}}  c@{\hspace{.8em}} 
c@{\hspace{.8em}} c@{\hspace{.8em}}} 
\hline \hline
        &      &      &      & SNIa &      &             \\
\hline 
Element & W7   & W70  & WDD1 & WDD2 & WDD3 & CDD1 & CDD2 \\
\hline
Si      & 0.16 & 0.14 & 0.27 & 0.21 & 0.16 & 0.28 & 0.20 \\
Fe      & 0.76 & 0.77 & 0.67 & 0.79 & 0.87 & 0.65 & 0.83 \\
Ni      & 0.13 & 0.10 & 0.04 & 0.06 & 0.07 & 0.04 & 0.06 \\
\hline
        &      &      &      &      &      &      &      \\

\end{tabular}
\begin{tabular}{  l@{\hspace{.8em}}  c@{\hspace{.6em}} 
c@{\hspace{.8em}} c@{\hspace{.8em}}} 
\hline
\hline
           &  & SNcc ~~~ ~~  &             \\
\hline
Element~~~ & Nomoto etal. (2006) & Chieffi\&Limongi (2004) & Woosley\&Weaver(1995) \\
\hline
   & Salpeter ~~~~top-heavy & Salpeter ~~~~top-heavy & Salpeter ~~~~top-heavy  \\
Si & 0.113 ~~~~~~~~~~ 0.124 & 0.163 ~~~~~~~~~~ 0.183 &  0.127 ~~~~~~~~~~ 0.132 \\
Fe & 0.090 ~~~~~~~~~~ 0.091 & 0.132 ~~~~~~~~~~ 0.135 &  0.109 ~~~~~~~~~~ 0.106 \\
Ni & 0.005 ~~~~~~~~~~ 0.005 & 0.014 ~~~~~~~~~~ 0.013 &  0.014 ~~~~~~~~~~ 0.013 \\
\hline
\hline
\end{tabular}

 \end{center}
 \end{table*}

 \begin{table*}
 \begin{center}
 \caption{Yields of Si and Ni relative to Fe in solar units of
 \citet{anders89} for the same SNe models used in
 Tab. \ref{tab_yields_SNe}}.
 \label{tab_ratios_SNe}
 \begin{tabular}{  l@{\hspace{.8em}}  c@{\hspace{.6em}} 
c@{\hspace{.8em}} c@{\hspace{.8em}}  c@{\hspace{.8em}}  
c@{\hspace{.8em}} c@{\hspace{.8em}}  c@{\hspace{.8em}} 
c@{\hspace{.8em}} c@{\hspace{.8em}}} 
\hline \hline
        &    &     &      & SNIa &      &       \\
\hline
Ratio  & W7   & W70  & WDD1 & WDD2 & WDD3 & CDD1 & CDD2 \\
\hline
Si/Fe  & 0.55 & 0.48 & 1.07 & 0.69 & 0.48 & 1.13 & 0.63 \\
Ni/Fe  & 4.20 & 3.23 & 1.42 & 1.85 & 2.09 & 1.35 & 1.85 \\
\hline
       &      &      &      &      &      &      &      \\
\end{tabular}
\begin{tabular}{  l@{\hspace{.8em}}  c@{\hspace{.6em}} 
c@{\hspace{.8em}} c@{\hspace{.8em}}} 
\hline
\hline
         &                     &SNcc~~~~~~~~~~           &             \\
\hline
Ratio~~~ & Nomoto etal. (2006) & Chieffi\&Limongi (2004) & Woosley\&Weaver(1995) \\
\hline
      & Salpeter ~~top-heavy & Salpeter ~~top-heavy  & Salpeter~~ top-heavy \\
Si/Fe & 3.28~~~~~~~  3.58 & 3.22 ~~~~~~~~  3.56 & 3.07 ~~~~~~~~  3.26 \\
Ni/Fe & 1.43~~~~~~~  1.38 & 2.71 ~~~~~~~~  2.43 & 3.24 ~~~~~~~~  2.99 \\
\hline
\end{tabular}

 \end{center}
 \end{table*}

The derived yields of Si, Fe and Ni in units of solar masses per
supernova explosion are reported in Table \ref{tab_yields_SNe}.  For
core-collapsed supernovae we report results for the standard Salpeter
and the AY top-heavy IMF.

Inspection of Table \ref{tab_yields_SNe} reveals how much yields
differ from one model to the other. Fe is characterized by the
smallest scatter, 10\% for SNIa and 18\% for SNcc, Si is less
constrained 27\% for SNIa and 19\% for SNcc, Ni features the largest
variations, 46\% for SNIa and 41\% for SNcc. Note also that the choice
of IMF produces only a modest impact on the yields.

The derived SNIa yields of Si and Ni relative to iron in
\citet{anders89} solar units are reported in Table
\ref{tab_ratios_SNe}.
Ratios in \citet{grevesse98} units can be easily computed by
multiplying both the Si/Fe and Ni/Fe values in Table
\ref{tab_ratios_SNe} by 0.675, whereas ratios in
\citet{lodders03} units are obtained by multiplying Si/Fe by 0.65 and
Ni/Fe by 0.68, and, ratios in \citet{asplund05} units by multiplying
Si/Fe by 0.67 and Ni/Fe by 0.72 (for the solar Ni we have used the
updated value by \citealp{scott09}).

The derived yields of Si and Ni relative to iron in solar units for
SNcc for different progenitor metallicities and IMFs are reported in
Table \ref{tab_ratios_SNe}. Conversion factors for different solar
units systems are the same as given above.

The yields relative to Fe for all our SNcc models are also listed in a
recent review paper \citep{werner08_rev}. Unfortunately some of the
values reported in that paper are incorrect, in some cases, by a large
amount, i.e. the Ni yields for the \citet{woosley95} models.
We have corresponded with the authors who are now considering
publishing a revised version of their tables.

\end{appendix}

\end{document}